\documentclass[sigconf,nonacm]{acmart}

\renewcommand\footnotetextcopyrightpermission[1]{}
\usepackage{algorithm}
\usepackage{algorithmic}
\usepackage{enumitem}
\usepackage{textcomp}
\usepackage{colortbl}
\usepackage{stmaryrd}
\usepackage{url}            
\usepackage{amsfonts}       
\usepackage{nicefrac}       
\usepackage{subfigure}
\usepackage{subcaption}
\usepackage{multirow}
\usepackage{makecell}

\newcommand{\methodname}{AEGIS }

\newcommand{\crpedit}[1]{{#1}}

\begin{document}

\title{Scaling Long-Sequence Homomorphic Encrypted Transformer Inference via Hybrid Parallelism on Multi-GPU Systems}


\author{Zhaoting Gong}
\email{zgong6@ncsu.edu}
\affiliation{%
  \institution{North Carolina State University}
  \city{Raleigh, NC}
  \country{USA}
}

\author{Ran Ran}
\email{rran@ncsu.edu}
\affiliation{%
  \institution{North Carolina State University}
  \city{Raleigh, NC}
  \country{USA}
}

\author{Fan Yao}
\email{fan.yao@ucf.edu}
\affiliation{%
  \institution{University of Central Florida}
  \city{Orlando, FL}
  \country{USA}
}

\author{Wujie Wen}
\email{wwen2@ncsu.edu}
\affiliation{%
  \institution{North Carolina State University}
  \city{Raleigh, NC}
  \country{USA}
}

\begin{abstract}
Fully Homomorphic Encryption (FHE) enables privacy-preserving Transformer inference, but long-sequence encrypted Transformers quickly exceed single-GPU memory capacity because encoded weights are already large and encrypted activations grow rapidly with sequence length. Multi-GPU execution therefore becomes unavoidable, yet scaling remains challenging because communication is jointly induced by application-level aggregation and encryption-level RNS coupling. Existing approaches either synchronize between devices frequently or replicate encrypted tensors across devices, leading to excessive communication and latency.

We present \methodname, an \underline{A}pplication-\underline{E}ncryption \underline{G}uided \underline{I}nference \underline{S}ystem for scalable long-sequence encrypted Transformer inference on multi-GPU platforms. \methodname derives device placement from ciphertext dependencies jointly induced by Transformer dataflow and CKKS polynomial coupling, co-locating modulus- and token-coherent data so that communication is introduced only when application dependencies require it, while reordering polynomial operators to overlap the remaining collectives with computation.

On  2048 tokens input, \methodname reduces inter-GPU communication by up to 57.9\% in FFNs and 81.3\% in self-attention versus prior state-of-the-art designs. On four GPUs, it achieves up to 96.62\% scaling efficiency, 3.86$\times$ end-to-end speedup, and 69.1\% per-device memory reduction. These results establish coordinated application-encryption parallelism as a practical foundation for scalable homomorphic Transformer inference.
\end{abstract}



\maketitle

\section{Introduction}

Machine Learning as a Service (MLaaS) has become a widely adopted paradigm that enables users to outsource model inference to cloud platforms equipped with large-scale GPU resources. While this paradigm offers convenient and scalable computation, it inherently exposes user inputs to the service provider during inference, raising significant privacy concerns due to potential data leakage. Homomorphic Encryption (HE)~\cite{gentry_fully_2009} addresses this issue by allowing arbitrary computations to be performed directly on encrypted data without decryption, thereby ensuring data confidentiality throughout the entire inference process. Several HE schemes based on the hardness of the Ring Learning With Errors (RLWE) problem, including BFV~\cite{fan_somewhat_2012}, BGV~\cite{brakerski_leveled_2014}, and CKKS~\cite{cheon_homomorphic_2016}, have been proposed to support encrypted arithmetic with different trade-offs between precision, noise growth, and efficiency. Despite their strong security guarantees, HE-based inference introduces substantial computational and memory overheads.

Existing work on HE-based private inference has largely focused on algorithmic optimizations to reduce single-device computation overhead, including ciphertext packing~\cite{kim_secure_2022, ran_cryptogcn_2022, zhang_secure_2025, lee_hetal_2024}, polynomial approximations for nonlinearities~\cite{park_powerformer_2024, lu_bumblebee_2023}, and compiler-level optimizations~\cite{dathathri_chet_2019,reagen_cheetah_2021,dathathri_eva_2020,lee_hecate_2022,lee_performance-aware_2024,cowan_porcupine_2021,cheon_dacapo_2024}. 
While these techniques achieve notable speedups, scaling HE inference to Transformer-based LLMs remains an open challenge due to the \emph{memory explosion} inherent to encrypted computation. For encrypted Transformer inference in particular, existing state-of-the-art systems that primarily target latency~\cite{zhang_secure_2025, pang_bolt_2023, moon_thor_2024} often omit detailed analysis of memory bottlenecks. As sequence length increases, both encoded model weights and ciphertext activations rapidly exceed the capacity of a single GPU, as illustrated in Figure~\ref{fig:memory-breakdown-by-sequence-length}, making multi-GPU execution unavoidable. Even for moderate workloads, encrypted BERT inference with 128 tokens already requires $4\times$ A100 GPUs~\cite{zhang_secure_2025}. 

This challenge is further amplified for long-context models. Under CKKS, weights and activations expand by orders of magnitude due to high-precision modulus chains (e.g., a 16-bit weight expands by approximately $110\times$ under a 1763-bit modulus~\cite{zhang_secure_2025}) and packing-induced redundancy. Consequently, even BERT-Base with a 2048-token context requires over 700\,GB of encoded weights and 110\,GB of ciphertext activations, exceeding the capacity of a single compute node and necessitating efficient multi-device inference. To handle large requests and reduce Time to First Token (TTFT) during prefill, plaintext LLM serving relies on multi-node parallel execution with a rich set of parallelization paradigms, most notably data parallelism (DP)~\cite{dean_large_2012} and tensor parallelism (TP)~\cite{narayanan_efficient_2021}, and their hybrid variants~\cite{qin_chimera_2025}, built atop efficient collective communication~\cite{using_1994_mpi}. However, new challenges arise in encrypted inference.

\textbf{First}, plaintext tensor parallelism is ineffective under encryption. These schemes fundamentally assume that (1) tensors can be arbitrarily partitioned and (2) collective communication efficiently synchronizes partial results across devices. Encrypted inference violates both assumptions: ciphertext packing encodes tokens into fixed slot layouts, preventing arbitrary tensor partitioning. As a result, directly adopting plaintext techniques either duplicates large encrypted activations that cause prohibitive memory overhead or yields redundant computation and communication due to slot-format constraints (Section~\ref{sec:exisiting-parallel-model-performs-bad}).

\textbf{Second}, HE exposes a native parallelization mechanism based on the \emph{Residue Number System (RNS)}~\cite{safavi-naini_homomorphic_2012}. Commonly referred to as \emph{limb parallelism} (see Figure~\ref{fig:rns-parallelization-model}), which is widely adopted in state-of-the-art HE accelerators~\cite{jayashankar_cinnamon_2025, al_badawi_multi-gpu_2021, wang_he-booster_2023}.
However, our profiling shows that this parallelism does not translate to scalable \emph{end-to-end} encrypted Transformer inference.
Despite evenly partitioned memory, performance is dominated by excessive inter-device communication during key switching~\cite{kim_accelerating_2023}, driven by frequent rotations in large and complex encrypted matrix multiplications (see Section~\ref{sec:analyzing-dependency-in-he-data-structures}).
On an RTX~A6000 system with NVLink, scaling to 2 GPUs even results in a \emph{2.25$\times$ slowdown}, demonstrating that it is fundamentally mismatched to Transformer workloads.

\textbf{Third}, managing collective communication itself is significantly more challenging than in plaintext. In encrypted inference, even a single token transmission requires sending multiple full ciphertexts, introducing extreme redundancy. For example, transmitting 1024 tokens requires only $\sim$2\,MB in plaintext but expands to approximately 1.3\,GB under encryption (see Table~\ref{tab:comm-plaintext-vs-encrypted}). This communication amplification dramatically increases latency and bandwidth demand, rendering conventional collective strategies ineffective.

\textbf{Together}, these challenges demonstrate that no existing solutions offer scalable encrypted Transformer inference, motivating a fundamentally different parallel execution model.
Existing systems fail to scale long-sequence encrypted Transformers because application-driven approaches are not encryption-aware, while encryption-driven approaches ignore application-level structure. To bridge this gap, we propose \methodname, a framework for scalable encrypted Transformer inference on multi-GPU platforms. \textbf{Our main contributions are as follows:}

\begin{itemize}[leftmargin=*]
    \item We present \textbf{\methodname}, a compiler-runtime co-designed framework for \textbf{scalable long-sequence encrypted Transformer inference} on multi-GPU platforms. It jointly orchestrates application-level data flow and low-level primitive polynomial execution to enable efficient multi-GPU parallelism.

    \item We propose a \textbf{dependency-aware workload partitioning strategy} that bridges application semantics and encryption constraints. \methodname incorporates both encryption-aware modulus-coherent placement and application-aware \emph{token-coherent placement} to co-locate slices of the same modulus chain or token, minimizing communication induced by application-level reductions.

    \item We develop a \textbf{polynomial-operator reordering mechanism} that restructures low-level instruction schedules to overlap collective communication with computation, thereby \textbf{hiding inter-device communication latency}. This design enables deterministic communication hiding without relying on runtime heuristics or dynamic synchronization.

    \item Experiments demonstrate that \methodname consistently outperforms state-of-the-art designs in both latency and memory efficiency, achieving up to \textbf{92.98\% scaling efficiency} and \textbf{1.86$\times$} end-to-end speed-up on two GPUs, while also reducing per-device memory consumption by up to \textbf{50.86\%} compared to baselines that replicate intermediate ciphertexts.
\end{itemize}

\section{Preliminaries}

\begin{table}
\centering
\caption{Notations and symbols used in this paper}
\label{tab:notations}
\vspace{-1em}
\resizebox{\columnwidth}{!}{%
\begin{tabular}{ll}
\toprule
\textbf{Notation} & \textbf{Description} \\ 
\addlinespace[2pt]\midrule
$N$                                & Polynomial degree (number of coefficients) \\
$S$                                & Number of available slots in an encoded message \\
$Q_L$                              & Ciphertext modulus chain $\{q_0, q_1, \dots, q_L\}$ \\
$P$                                & Special modulus chain $\{q_{L+1}, q_{L+2}, \dots\}$ \\
$L$                                & Length of the modulus chain (maximum multiplicative depth) \\
$l \in [0, L)$                     & Current remaining multiplicative level \\
\addlinespace[2pt]\midrule
$G$                             & Number of GPU devices \\
\addlinespace[2pt]\midrule
$d_{\text{model}}$              & Embedding dimension of Transformer layers \\
$d_{\text{head}}$               & Dimensionality of each attention head \\
$T$                             & Input sequence length (number of tokens) \\
\addlinespace[2pt]\midrule
$s_{\text{tok}}$                & Number of elements of each token encoded into one ciphertext \\
$c_{\text{tok}}$                & Number of ciphertexts used to represent a single token \\
\bottomrule
\end{tabular}%
}
\vspace{-0.5em}
\end{table}

\subsection{Homomorphic Encryption and Operations}

RLWE-based HE schemes perform computation over the polynomial residue ring $\mathbb{Z}_q[x]/(x^N + 1)$, where $N$ (a power of two) is the ring dimension and $q$ the ciphertext modulus. In CKKS, a modulus chain $Q = \prod_{i=0}^{L} q_i$ is used to support multiple levels of computation. A plaintext vector of length $S = N/2$ is encoded as a complex polynomial and encrypted under a secret key, where $S$ denotes the \emph{slot number}, which means the number of complex values that can be packed into one ciphertext.  
The packed values are processed simultaneously, enabling Single Instruction Multiple Data (SIMD)-like computation across slots in subsequent homomorphic operations.

CKKS supports linear operations \texttt{PAdd}, \texttt{CAdd}, \texttt{PMult}, and \texttt{CMult}, each multiplication increasing both scale and noise and thereby consuming one level of the modulus chain.  
The \texttt{Rescale} operation divides by the last modulus $q_L$ to reduce scale and remove one prime, while \texttt{Relinearization} (\texttt{Relin}) restores ciphertext size and suppresses noise growth through key switching.  
Additional primitives include rotations \texttt{Rot}$_\rho$, which cyclically shift plaintext slots by offset $\rho$, and bootstrapping~\cite{bossuat_efficient_2021, chen_improved_2018, jung_over_2021} (\texttt{Boot}), which replenishes the modulus chain to enable deeper computation.

Scalar operations (\texttt{P/CAdd}, \texttt{P/CMult}) act only on the main primes $(q_0, \dots, q_L)$, whereas key-switching operations (\texttt{Relin}, \texttt{Rot}, \texttt{Boot}) temporarily extend the modulus to $\mathbb{Z}_{Q \cdot P}$, where $P$ is the product of \emph{special primes} used for digit decomposition.  
After switching, ciphertexts are reduced back to $\mathbb{Z}_Q$, and the special primes remain active only during key-switching.

\subsection{Data Structure in Homomorphic Encryption}

\begin{figure}
    \centering
    \includegraphics[width=\linewidth]{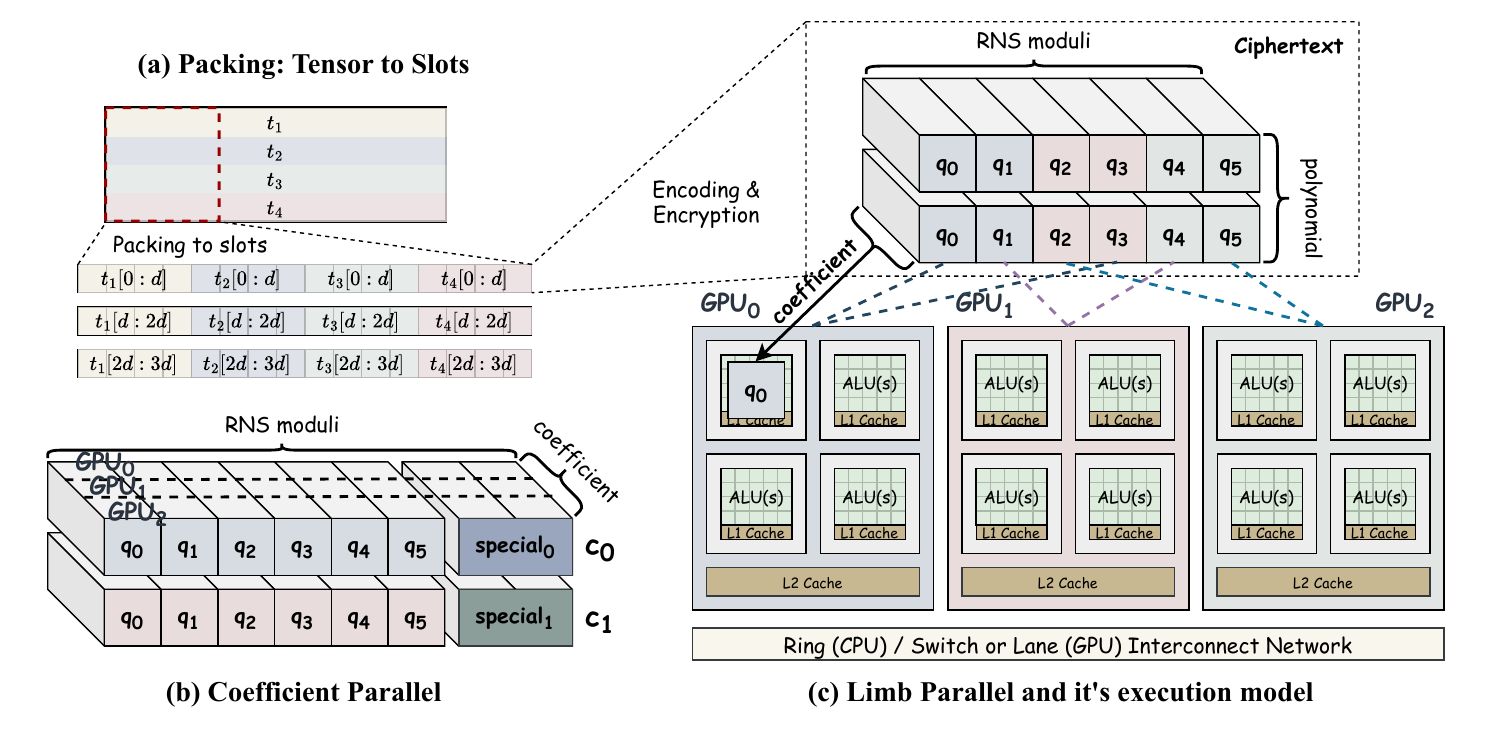}
    \caption{Illustration of data layout and parallelization models in CKKS computation.  
    (a)~Packing that maps plaintext tensors into ciphertext slots.  
    (b)~Coefficient-parallel, distributing work across polynomial coefficients.  
    (c)~Limb-parallel, distributing work across RNS primes and exposing thread parallel on coefficient.}
    \Description{Three-part illustration of CKKS data layout and parallelization. It shows plaintext values packed into ciphertext slots, coefficient-parallel execution across polynomial coefficients, and limb-parallel execution across RNS primes.}
    \label{fig:rns-parallelization-model}
    \vspace{-1em}
\end{figure}

SOTA CKKS frameworks~\cite{lattigo, sealcrypto, OpenFHE, Liberate_FHE} employ the Chinese Remainder Theorem (CRT) to represent large polynomial coefficients in Residue Number System (RNS) form, enabling efficient arithmetic under large moduli.  
Each ciphertext component is decomposed over primes $(q_0, q_1, \dots, q_L)$, where every coefficient $c \in \mathbb{Z}_Q$ is stored as residues $\{\, c \bmod q_i \,\}$.  
The remaining number of primes determines the ciphertext’s \emph{available multiplicative depth}: fewer primes yield smaller ciphertexts and reduced capacity.

As illustrated in Figure~\ref{fig:rns-parallelization-model}, \textbf{all data objects in CKKS including plaintexts, ciphertexts, and keys, share a common algebraic basis $\mathcal{R}_Q$ and can be viewed as structured tensors.}  
The coefficient and modulus dimensions correspond to the polynomial degree $N$ and the modulus chain $Q$. A ciphertext $ct = (c_0, c_1) \in \mathcal{R}_Q^2$ contains two polynomial components with $N$ coefficients replicated across all primes in $Q$.  
Conceptually, it forms a 3D tensor with RNS, coefficient, and polynomial dimensions. 
GPU implementations typically adopt an RNS-major layout, storing residues under the same modulus contiguously across coefficients, improving coalesced global-memory access, enabling per-prime kernel launches, and aligning with RNS-partitioned execution.

\subsection{Parallel Residual Number System}\label{sec:parallelism-in-residual-number-system}

Since every encrypted tensor in CKKS is structured within the RNS, multi-GPU execution can be viewed as parallelizing the RNS tensor across devices.  
Two RNS-native strategies are used: \emph{coefficient partitioning}~\cite{al_badawi_multi-gpu_2021, wang_he-booster_2023}, which distributes polynomial coefficients among GPUs (Figure~\ref{fig:rns-parallelization-model}(b)), and \emph{limb partitioning}, which assigns subsets of primes in the modulus chain to different devices~\cite{fan_tensorfhe_2023, jayashankar_cinnamon_2025, kim_bts_2022, wang_he-booster_2023, deng_trinity_2024}.  
With typical CKKS parameters containing multiple primes $q_0, \dots, q_{L-1}$, limb partitioning achieves scalable parallelism by distributing prime-indexed tensors across devices (Figure~\ref{fig:rns-parallelization-model}(c)), while each device exposes coefficient-level parallelism through its streaming multiprocessors (SMs).

Formally, device $\text{GPU}_d$ processes coefficient tensors $\{\, q_i \mid i \bmod D = d \,\}$, where $D$ is the number of devices.  
For example, in Figure~\ref{fig:rns-parallelization-model}(c), a ciphertext with $Q=\{q_0, q_1, \dots, q_5\}$ is mapped onto three GPUs: $\text{GPU}_0 \!\rightarrow\! \{q_0, q_3\}$, $\text{GPU}_1 \!\rightarrow\! \{q_1, q_4\}$, and $\text{GPU}_2 \!\rightarrow\! \{q_2, q_5\}$. \textbf{RNS parallelization is inherently application-independent}: it operates directly on the ciphertext’s modular structure and partitions work according to the modulus chain, irrespective of how the plaintext is encoded.

\begin{figure}
    \centering
    \includegraphics[width=0.8\linewidth]{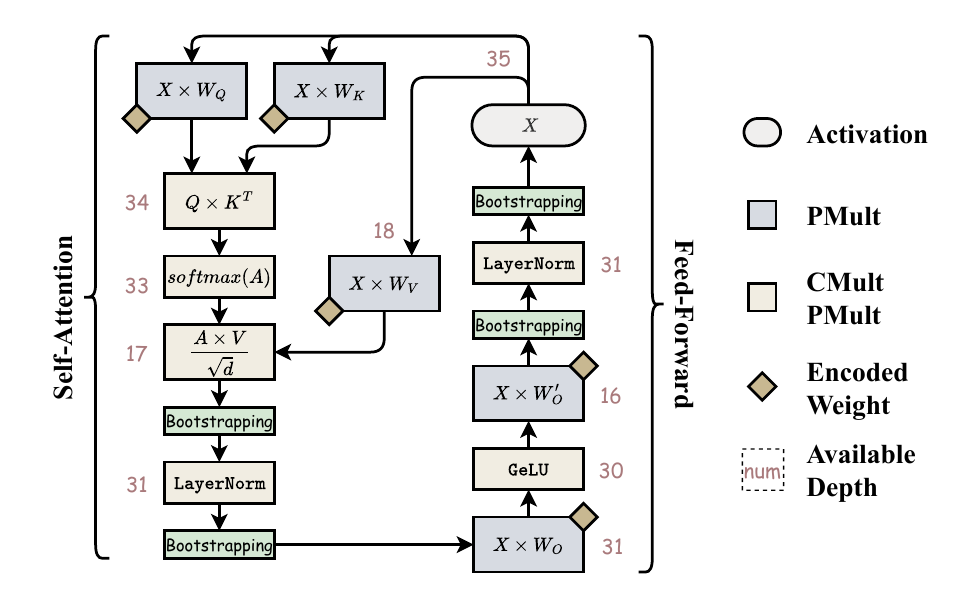}
    \caption{Encrypted Transformer inference pipeline with bootstrapping placement (NEXUS~\cite{zhang_secure_2025}).}
    \Description{Pipeline diagram of encrypted Transformer inference, showing linear layers, attention, nonlinear layers, and where bootstrapping is inserted to restore multiplicative depth.}
    \label{fig:encryoted-transformer-inference-diagram}
\vspace{-1em}
\end{figure}

\subsection{Encrypted Transformer Inference}

A Transformer block consists of self-attention and feed-forward layers.
Given an input tensor $X$, linear projections produce $Q$, $K$, and $V$, followed by attention computation $A=\texttt{softmax}(QK^\top/\sqrt{d})$, aggregation $AV$, and a feed-forward network with activation, normalization, and residual connections.
During the prefill phase, inference processes sequences of up to thousands of tokens, and runtime memory is dominated by model parameters and activations.

In encrypted inference, the client encrypts $X$ into ciphertexts and sends them to the server for evaluation.
As shown in Figure~\ref{fig:encryoted-transformer-inference-diagram}, Transformer operations are mapped to homomorphic primitives: linear layers use \texttt{PMult}, attention and activations use \texttt{CMult}, and nonlinear functions are approximated by high-degree polynomials combining both.
Since each multiplication consumes one modulus level, bootstrapping (\texttt{Boot}) is periodically inserted to restore noise budget and enable deep Transformer execution.

Encrypted inference incurs substantial memory overhead from three sources: ciphertext activations, key-switching keys (for rotations, \texttt{CMult}, and bootstrapping), and pre-encoded plaintext weights.
Pre-encoding weights is essential, as it trades storage for reduced computation. For example, reducing a BERT linear projection over 128 tokens from $354.31\,\mathrm{s}$ to $270.22\,\mathrm{s}$.

\begin{figure}
    \centering
    \includegraphics[width=0.9\linewidth]{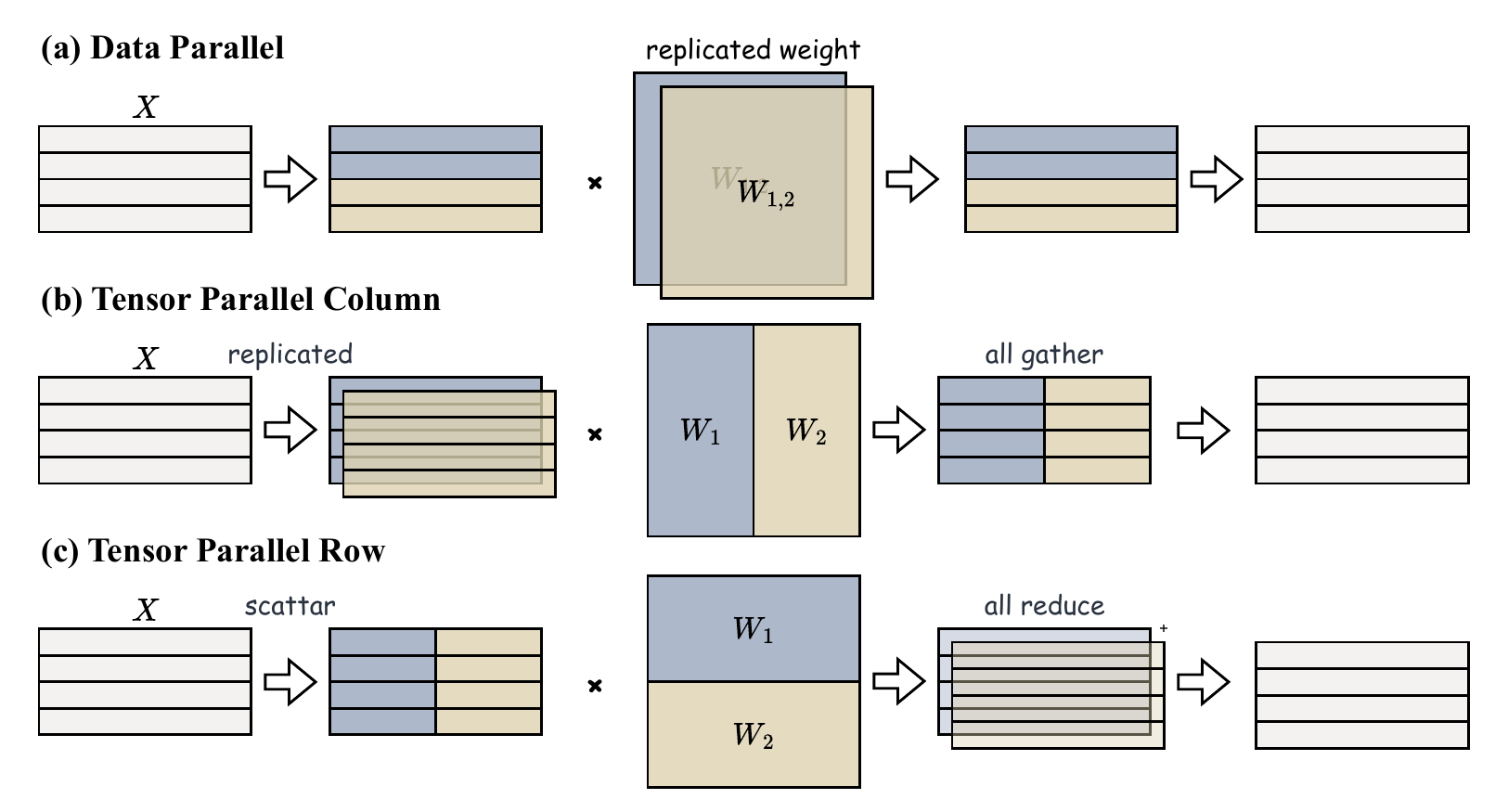}
    \caption{Fundamental parallelisms in plaintext inference.
    (a)~Data parallelism replicates model weights across devices.
    (b,c)~Tensor parallelism partitions weights and aggregates partial results via collectives.}
    \Description{Three-panel overview of plaintext inference parallelism. It contrasts data parallelism, which replicates weights, with tensor parallelism, which shards weights and then uses collectives to combine partial results.}
    \label{fig:parallelisms-in-plaintext-inference}
    \vspace{-1em}
\end{figure}

\subsection{Parallel Models and Collective Primitives}

Plaintext LLM serving relies on hybrid parallelism~\cite{shoeybi_megatron-lm_2020, lepikhin_gshard_2020, li_sequence_2023, pope_efficiently_2023} to scale model capacity and reduce time-to-first-token (TTFT).
On multi-GPU systems, these approaches are built upon fundamental partitioning schemes, data parallelism (DP)~\cite{dean_large_2012} and tensor parallelism (TP)~\cite{narayanan_efficient_2021}, as illustrated in Figure~\ref{fig:parallelisms-in-plaintext-inference}.
Each GPU processes a shard of the tensor and periodically exchanges intermediate results, trading computation for inter-GPU communication over interconnects such as NVLink or PCIe.

This communication is supported by collective primitives such as \texttt{Send\&Recv}, \texttt{AllReduce}, \texttt{AllGather}, and \texttt{AllToAll}, which enable scalable synchronization without explicit point-to-point orchestration.
For example, \texttt{AllReduce} synchronizes gradients in DP~\cite{dean_large_2012}, while \texttt{AllGather} aggregates partial activations in TP~\cite{shoeybi_megatron-lm_2020}.
Recent systems further reduce communication overhead by exploiting structured dependencies in model architectures~\cite{narayanan_efficient_2021, fedus2022switch}.

Similarly, GPU-accelerated CKKS inference employs inter-device communication to coordinate encrypted polynomial fragments partitioned across coefficients or RNS limbs.
Prior work uses explicit \texttt{Send\&Recv} to convert between coefficient- and limb-parallel layouts~\cite{al_badawi_multi-gpu_2021}, or enforces hard synchronization via \texttt{AllGather} to reconcile modulus-dependent ciphertext components~\cite{Liberate_FHE}.

\subsection{Threat Model and Security Level}

We assume a \emph{semi-honest} threat model, in which the cloud server follows the prescribed inference protocol but may attempt to infer information from encrypted data.  
The client encrypts inputs locally, uploads ciphertexts and public keys to an untrusted MLaaS provider, and decrypts the returned results.

Security under CKKS relies on the hardness of the Ring Learning With Errors (RLWE) problem at a target security level $\lambda$ (typically 128 bits).  
The parameter pair $\{N, Q\}$ jointly determines ciphertext precision and size.  
A larger modulus chain $Q=\{q_0,\dots,q_L\}$ supports greater multiplicative depth, which is required by deeper models (approximately $5$--$8$ for CNNs, $10$--$12$ for GCNs, and over $20$ for Transformers), but also increases ciphertext size, as storage grows with both $\lvert Q\rvert$ and the bit-length of each $q_i$. For Transformer, this typically results in a modulus chain exceeding 1600 bits.

\section{Motivation}\label{sec:motivation}

\begin{figure}
    \centering
    \includegraphics[width=0.9\linewidth]{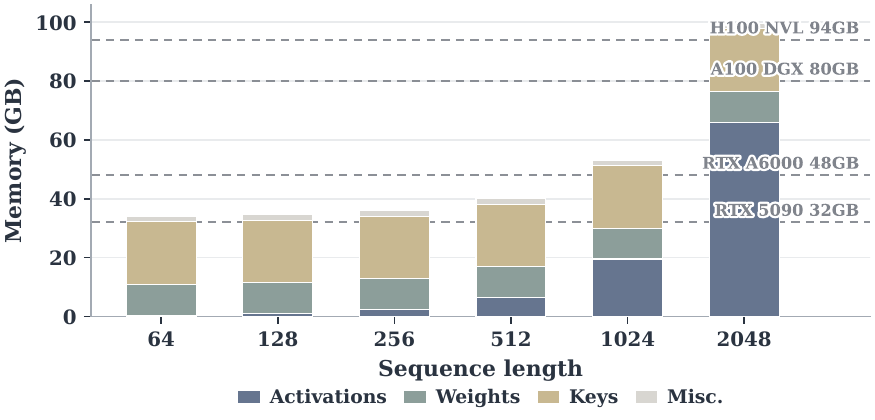}
    \caption{Memory breakdown of encrypted BERT-Base inference across input sequence lengths. The stacked bars represent the major memory components. Common NVIDIA GPU memory capacities are marked for reference.}
    \Description{Stacked bar chart of encrypted BERT-Base memory versus sequence length. Activation memory grows rapidly and becomes dominant at long sequences, exceeding common GPU memory capacities.}
    \label{fig:memory-breakdown-by-sequence-length}
    \vspace{-1em}
\end{figure}

\subsection{Distributing Large-Scale HE Inference}
\label{sec:distributed-large-scale-he-inference}

Encrypted Transformer inference incurs substantial memory overhead from pre-encoded weights, ciphertext activations, and user keys.
Although weights can be streamed layer-by-layer from CPU memory, even moderate-scale models expand to hundreds of gigabytes after encoding.
Ciphertext activations grow rapidly with sequence length and soon dominate runtime memory usage, as shown in Figure~\ref{fig:memory-breakdown-by-sequence-length}.
At sequence length 128, runtime memory already includes $666.89\,\mathrm{MB}$ of activations, $10.64\,\mathrm{GB}$ of encoded weights, and $20.45\,\mathrm{GB}$ of user keys.
At sequence length 2048, activations alone reach $52.95\,\mathrm{GB}$, exceeding the capacity of a single GPU even for a small model like BERT-Base~\cite{devlin_bert_2019}.
Consequently, practical encrypted inference \emph{necessarily} requires distributing both ciphertexts and weights across multiple devices.

However, efficient distribution is challenging because encrypted computation induces communication patterns fundamentally different from plaintext inference.
At the application level, cross-device reductions occur along either the token or embedding dimension, but in the encrypted domain these reductions require transmitting entire ciphertexts. Without loss of generality, under a packing layout $(s_{\text{tok}}, c_{\text{tok}})$, each ciphertext allocates $s_{\text{tok}}$ slots to one token, so a $d_{\text{model}}$- dimensional embedding is first split across $c_{\text{tok}}=\lceil d_{\text{model}}/s_{\text{tok}}\rceil$ ciphertexts. Each ciphertext packs multiple tokens horizontally, up to $S/s_{\text{tok}}$ tokens per ciphertext, where $S$ is the total slot count.
Transmitting $T$ tokens therefore requires
$K_{\text{send}}(T)=\lceil T/(S/s_{\text{tok}})\rceil\cdot c_{\text{tok}}$
ciphertexts at level $l$, yielding a point-to-point communication volume of
$\text{P2P}(T,l)\approx K_{\text{send}}(T)\cdot (2\cdot l\cdot 2N\cdot 8~\mathrm{bytes})$.
Collective operations amplify this cost:
$\text{AllReduce}(T,l)\approx (G-1)\cdot \text{P2P}(T,l)$ and
$\text{AllGather}(T,l)\approx (G-1)\cdot \text{P2P}(T/G,l)$.

This highlights the severe communication amplification inherent to encrypted inference.
Limb-parallel designs are constrained by RNS partitioning and frequent key switching, while tensor-parallel designs scale with both $c_{\text{tok}}$ and the multiplicative depth $l$. For example, transmitting 1024 tokens at full depth on two devices requires 1.3\,GB for P2P transfer, 3.8\,GB for \texttt{AllReduce}, and 0.96\,GB for \texttt{AllGather}, compared to only a few megabytes in plaintext (Table~\ref{tab:comm-plaintext-vs-encrypted}).
These costs demonstrate that while distributing HE inference is unavoidable, achieving scalability critically depends on minimizing when, where, and how collectives are invoked.

\begin{table*}
\centering
\vspace{-1em}
\caption{
Per-device communication cost of common collectives in plaintext and encrypted inference. Example assumes $T{=}1024$, $d_{\text{model}}{=}1024$, bf16, full slot utilization, $S{=}32768$, and bootstrappable CKKS parameters ($\sim$40\,MB ciphertexts at $l{=}35$). Encrypted communication is 2--3 orders of magnitude more expensive than plaintext.}
\label{tab:comm-plaintext-vs-encrypted}
\vspace{-1em}
\renewcommand{\arraystretch}{1.2}
\resizebox{0.9\textwidth}{!}{
\begin{tabular}{lccccc}
\toprule
\textbf{Collective} 
& \textbf{Plaintext cost} 
& \textbf{Plaintext example} 
& \textbf{Encrypted cost} 
& \textbf{Encrypted example} \\
\midrule

\texttt{P2P} send $(T$ tokens$)$ 
& $T \cdot d_{\text{model}} \cdot \texttt{sizeof(bf16)}$
& $1024 \cdot 1024 \cdot 2 \approx 2$\,MB
& $K_{\text{send}}(T) \cdot (2 \cdot l \cdot 2N \cdot 8)$
& $32 \times 40$\,MB $\approx 1.3$\,GB \\

\texttt{AllReduce} $(T$ tokens$)$ 
& $(G-1)\,T \cdot d_{\text{model}} \cdot \texttt{sizeof(bf16)}$
& $3 \times 2$\,MB $= 6$\,MB
& $(G-1)\,K_{\text{send}}(T) \cdot (2 \cdot l \cdot 2N \cdot 8)$
& $3 \times 1.3$\,GB $\approx 3.8$\,GB \\

\texttt{AllGather} $(T$ tokens$)$ 
& $(G-1)\,\frac{T}{G} \cdot d_{\text{model}} \cdot \texttt{sizeof(bf16)}$
& $3 \times 0.5$\,MB $= 1.5$\,MB
& $(G-1)\,K_{\text{send}}\!\left(\frac{T}{G}\right) \cdot (2 \cdot l \cdot 2N \cdot 8)$
& $3 \times (8 \times 40$\,MB$) \approx 0.96$\,GB \\

\bottomrule
\end{tabular}}
\end{table*}

\subsection{Limitations of Existing Parallel Models}

\label{sec:exisiting-parallel-model-performs-bad}

\subsubsection{Data Dependency in HE-based Inference}
\label{sec:analyzing-dependency-in-he-data-structures}

Encrypted inference applications are expressed as HE-operator control flows and compiled into polynomial instructions. \textbf{Each abstraction layer---application, HE operator, and polynomial instruction---exhibits distinct yet interrelated data dependencies, together forming a complex multi-dimensional coupling structure that challenges efficient parallelization.}

At the \emph{application level}, model semantics define structured dependencies among ciphertexts: \texttt{Linear} layers perform embedding-wise aggregation, attention blocks perform head-wise aggregation, and normalization layers perform token-wise aggregation.  
These patterns, realized through ciphertext packing, propagate coupling along both coefficient and modulus dimensions and directly influence pipeline organization.

At the \emph{HE operator level}, HE primitives introduce intrinsic coupling: \texttt{Mult} induces coefficient-wise dependencies through the \texttt{NTT} / \texttt{INTT} pipeline, while \texttt{Rot} introduces both coefficient- and modulus-wise dependencies due to \texttt{KeySwitch} within it.  
At the \emph{polynomial instruction level}, ciphertexts exhibit structured dependencies across coefficients and modulus chains. Operations such as \texttt{NTT}/\texttt{INTT} and automorphisms create intra-prime coupling, where each output coefficient depends on all input coefficients within the same modulus, while cross-prime coupling arises during residue reconstruction and base conversion in \texttt{KeySwitch}, where new residues are computed from all lower-level residues in \texttt{ModUp}.

\subsubsection{Limb Parallelism Is Communication-Bound and Depth-Limited}

Profiling inter-device traffic shows that \emph{key switching} dominates cross-device communication in encrypted Transformer workloads. Rotations in matrix multiplications, polynomial nonlinearities, and relinearization all induce instruction-level coupling across RNS limbs of the same ciphertext component, making repeated synchronization unavoidable under limb-parallel execution.

\begin{figure}
    \centering
    \includegraphics[width=0.9\linewidth]{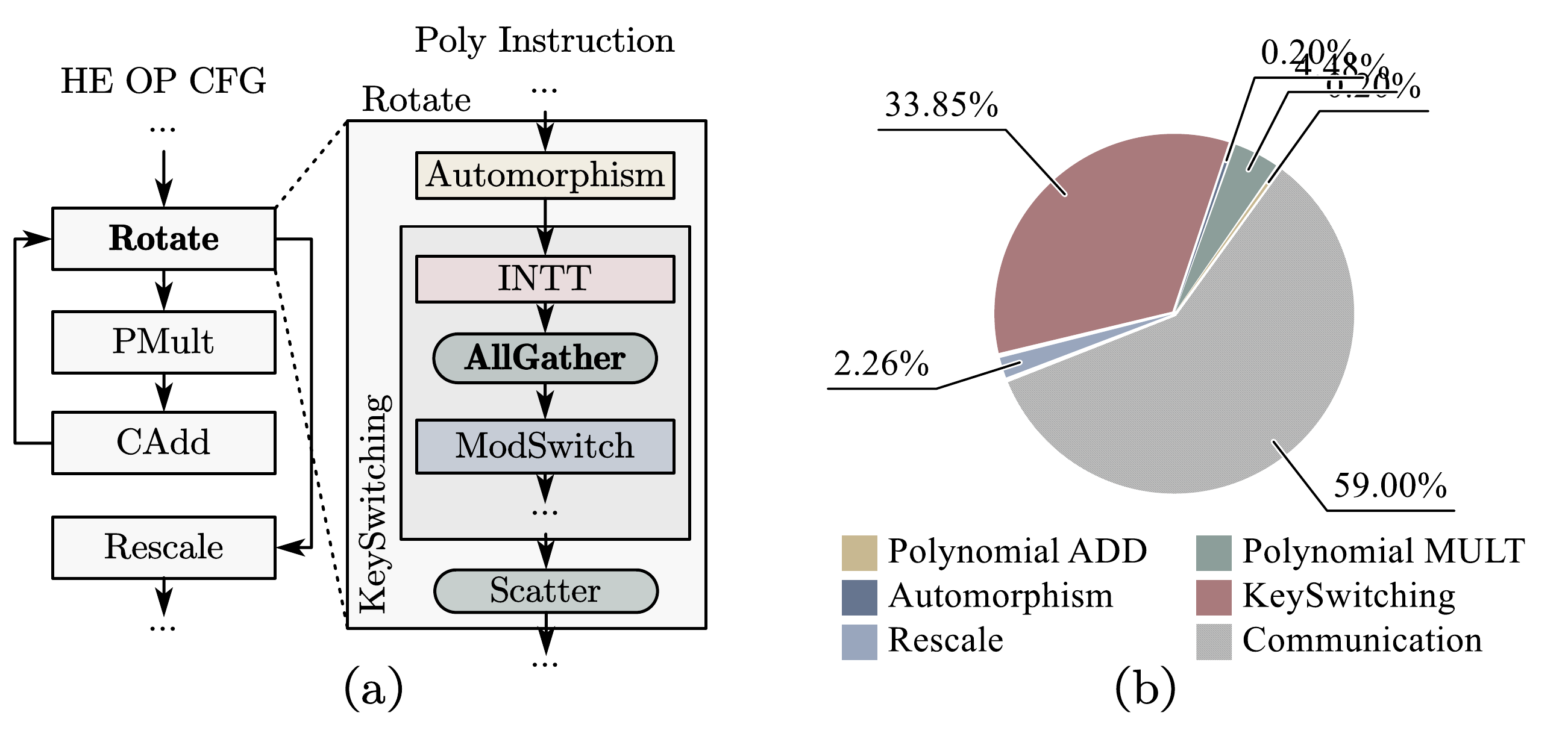}
    \caption{(a) Collective in RNS-parallel BSGS matrix multiplication down to compute graph level.
    (b) Latency breakdown of ciphertext matrix multiplication on a 2-GPU with NVLink.}
    \Description{Two-panel figure. Panel a shows collectives inserted repeatedly during RNS-parallel BSGS ciphertext matrix multiplication. Panel b shows that on two GPUs with NVLink, communication and key switching dominate ciphertext matrix multiplication latency.}
    \label{fig:limb-ccmult-mm-latency-breakdown}
    \vspace{-1em}
\end{figure}

As a result, limb parallelism scales poorly, even negatively across GPUs. On an RTX~A6000 system, a $512\times512$ ciphertext matrix multiplication takes $16.7\,\mathrm{s}$ on one GPU, but increases to $37.5\,\mathrm{s}$ on two GPUs and $62.3\,\mathrm{s}$ on four GPUs.  
In the two-GPU case, communication alone accounts for $59\%$ of runtime, with key switching contributing an additional $34\%$. Down to the computation graph, in BSGS matrix multiplication~\cite{kim_secure_2022, ebel_orion_2025}, each modulus-switching step triggers an \texttt{AllGather} to synchronize RNS limbs across devices (Figure~\ref{fig:limb-ccmult-mm-latency-breakdown}(a)). The total rotation cost accumulates as $t_{\mathrm{rot}} = n_{\mathrm{rot}} \cdot (\tau_{\mathrm{ks}} + \tau_{\mathrm{comm}})$, and long-sequence inference requires tens of thousands of rotations~\cite{kim_secure_2022, ran_penguin_2023, juvekar_gazelle_2018, pang_bolt_2023}. Consequently, communication dominates latency; Figure~\ref{fig:limb-ccmult-mm-latency-breakdown}(b) shows that $59\%$ of a single matrix multiplication is spent waiting on communication.

Beyond communication costs, limb parallelism is fundamentally limited by modulus depth.  
A level-$l$ ciphertext can be partitioned across at most $l$ devices, since each RNS limb corresponds to one modulus prime.  
Thus, a ciphertext with $l=6$ cannot scale beyond six GPUs regardless of tensor size or sequence length (Figure~\ref{fig:rns-parallelization-model}).  
In practice, even bootstrappable workloads with $l\!\approx\!14$~\cite{zhang_secure_2025, moon_thor_2024} are insufficient for scalable long-sequence inference (see Section~\ref{sec:scaling-analysis}).

\subsubsection{Conventional Tensor Parallelism Suffers Redundancy}

\begin{figure}
    \centering
    \includegraphics[width=\linewidth]{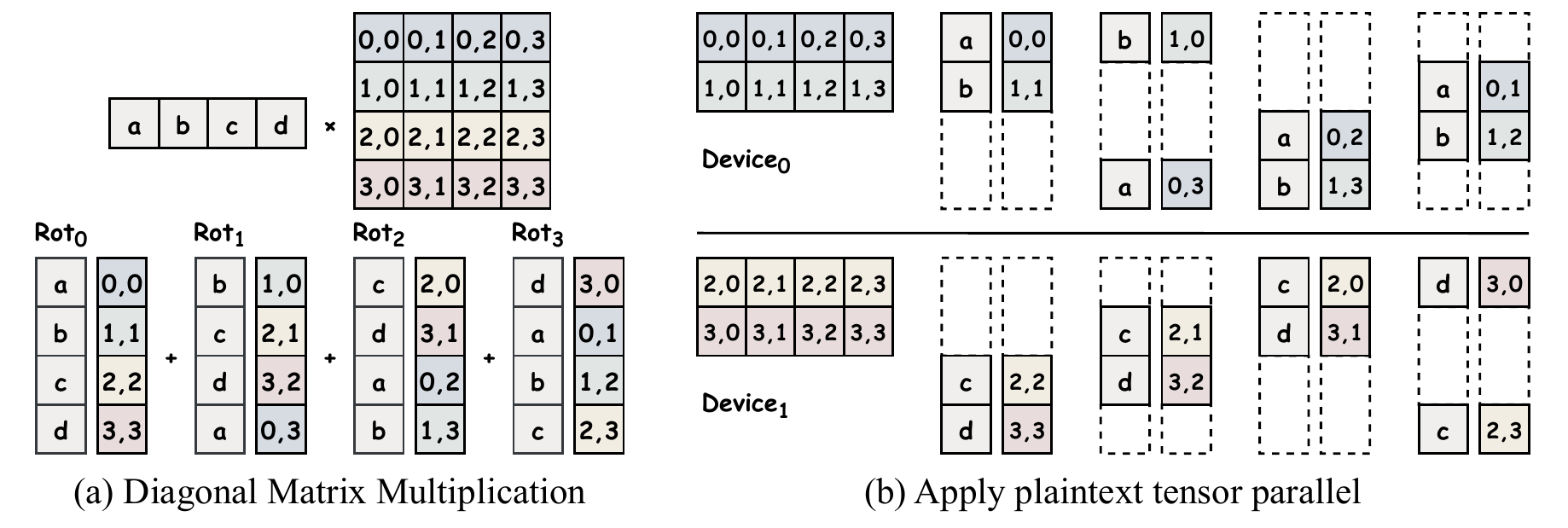}
    \caption{(a) Encrypted diagonal matrix multiplication.
    (b) Directly applying plaintext tensor parallelism induces redundant ciphertext replication.}
    \Description{Two-panel diagram. Panel a illustrates encrypted diagonal matrix multiplication. Panel b shows that directly reusing plaintext tensor parallelism replicates ciphertexts across devices and causes redundant encrypted work.}
    \label{fig:directly-apply-plaintext-tensor-parallel-cause-redundancy}
    \vspace{-1em}
\end{figure}

Plaintext-style tensor parallelism (TP) introduces severe inefficiencies when directly applied to homomorphic encrypted inference.
Under typical CKKS packing, each token is represented by
$c_{\text{tok}} = \left\lceil \tfrac{d_{\text{model}}}{s_{\text{tok}}} \right\rceil$
ciphertexts.
Column tensor parallelism (Figure~\ref{fig:parallelisms-in-plaintext-inference}(b)) replicates encrypted inputs across devices, quickly exhausting GPU memory since ciphertext tensors are orders of magnitude larger than their plaintext counterparts.
Row tensor parallelism (Figure~\ref{fig:parallelisms-in-plaintext-inference}(c)) fares no better: partitioning along the embedding dimension still requires each device to hold all $c_{\text{tok}}$ ciphertexts to reconstruct complete token features, resulting in redundant storage and duplicated computation.
This inefficiency is illustrated in Figure~\ref{fig:directly-apply-plaintext-tensor-parallel-cause-redundancy}.

Furthermore, because each ciphertext decomposes into per-prime polynomial fragments, transferring even a single token requires communicating all $c_{\text{tok}}$ ciphertexts across every modulus in $Q_L$. This inflates communication volume by orders of magnitude relative to plaintext execution, rendering plaintext-style token-aware collectives fundamentally inefficient under HE.

\section{\methodname}

As outlined in Section~\ref{sec:motivation}, supporting long input sequences on a multi-GPU system exposes two primary bottlenecks: memory usage and communication overhead. To address them, a parallelization strategy that jointly considers \emph{application-aware} and \emph{encryption-aware} optimizations is required, since data dependencies arise simultaneously from Transformer computation patterns and RNS-level HE operations. In this section, we introduce \methodname, a multi-GPU execution framework designed to achieve high-throughput for long-sequence encrypted Transformer inference through coordinated optimization across the application layer and the polynomial-instruction layer. We begin by presenting a \emph{polynomial placement algorithm} that assigns ciphertext components and HE operators to devices through a unified analysis of application-level data dependencies and RNS-level coupling among polynomial limbs. To further alleviate communication overhead, we introduce a \emph{polynomial–operator reordering} mechanism that restructures low-level instruction schedules to overlap collective communication with computation, effectively hiding inter-device latency.

\subsection{Dependency-Aware Workload Partitioning}
\label{sec:application-aware-polynomial-partitioning}

The fundamental mismatch between \emph{application-level} tensor semantics and the
\emph{encryption-level} Residue Number System (RNS) representation in CKKS arises from how
existing HE inference systems lower high-level application graphs into polynomial instructions. At runtime, however, each device observes only local polynomial fragments, either coefficient or modulus (prime) slices, without visibility into token boundaries, packing layouts, or aggregation semantics. As a result, plaintext-style partitioning and collectives require full ciphertext reconstruction, inducing excessive cross-device communication and yielding scalability that is dominated by RNS coupling rather than problem size.

The key observation is that distributed CKKS execution is dominated by \emph{reduction operations} arising from both RNS coupling and application-level aggregation, including embedding-wise accumulations in linear projections, sequence-wise reductions in attention, and reductions in layer normalization and softmax.
To address this, \methodname adopts a \textbf{communication-efficient, encryption-aware polynomial partitioning strategy} that jointly considers RNS dependencies and Transformer aggregation patterns, prioritizing co-location--aware data placement as introduced below.

\textbf{Modulus-coherent placement.}
Conventional limb-parallel schemes assign polynomials independently by global index
$\{\, p \mid p_i,\; i \bmod G = j \,\} \rightarrow \text{device } j$,
treating each polynomial as an isolated unit and ignoring cross-limb dependencies induced by key switching and rotations.
In contrast, \methodname enforces modulus-coherent placement through a \emph{polynomial-level dependency analysis} of the HE computation graph.
After lowering the application graph into polynomial operators, \methodname identifies groups of polynomials that must participate jointly in key switching or rotation chains.
All polynomials within such a dependency group are first mapped to the same device before the next key-switching operation in the computation graph, ensuring that key switching and rotation remain local and eliminating cross-device synchronization.
When the number of dependency groups is not divisible by $G$, complete groups are assigned first, and splitting is used only as a fallback for load balancing with communication hiding (see Section~\ref{sec:communication-hiding}).

\textbf{Token-coherent placement.}
\methodname additionally performs an \emph{application-level analysis} before polynomial lowering to identify token-wise dependencies.
Using the packing layout $(s_{\text{tok}}, c_{\text{tok}})$ introduced in Section~\ref{sec:distributed-large-scale-he-inference}, each token spans $c_{\text{tok}}$ ciphertexts.
For $c_{\text{tok}}>1$, dependency groups corresponding to different slices of the same token are co-located whenever possible.
This token-coherent placement aligns computation with embedding-wise reductions and avoids synchronizing partial token results across devices.
\crpedit{Although we do not focus on packing design in this paper, the placement rule itself is packing-agnostic: when $(s_{\text{tok}}, c_{\text{tok}})$ changes, \methodname re-derives ciphertext dependency groups from the new layout and applies the same modulus-first, token-second hierarchy.}
Together, \methodname\ follows a strict placement hierarchy in which modulus chain dependency is of first priority and token coherence is secondary priority; these principles underpin the workload partitioning strategies in Section~\ref{sec:workload-partition-of-ffns} and
Section~\ref{sec:workload-partition-of-self-attn}.

\begin{figure}
    \centering
    \includegraphics[width=\linewidth]{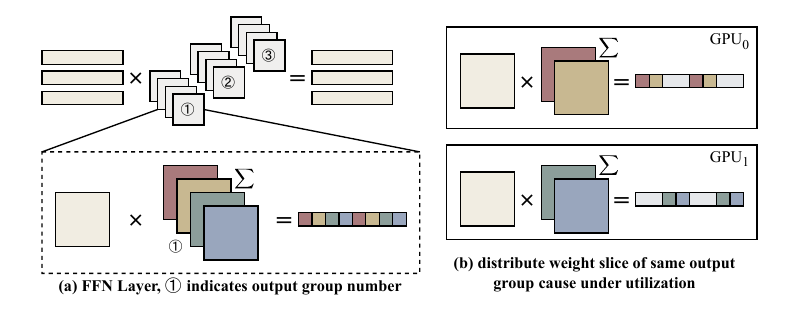}
    \caption{Output-slot coupling: (a) weight slices within the same output group generate results on the same slots; (b) splitting token output-coupled weight onto 2 devices leads to half slot waste in the next computation.}
    \Description{Two-panel diagram of output-slot coupling in encrypted linear layers. Weight slices for the same output group map to the same slots, and splitting them across two devices leaves many output slots unused in the next step.}
    \label{fig:weight-slice-output-dependency}
    \vspace{-1em}
\end{figure}

\begin{figure*}
    \centering
    \includegraphics[width=0.95\linewidth]{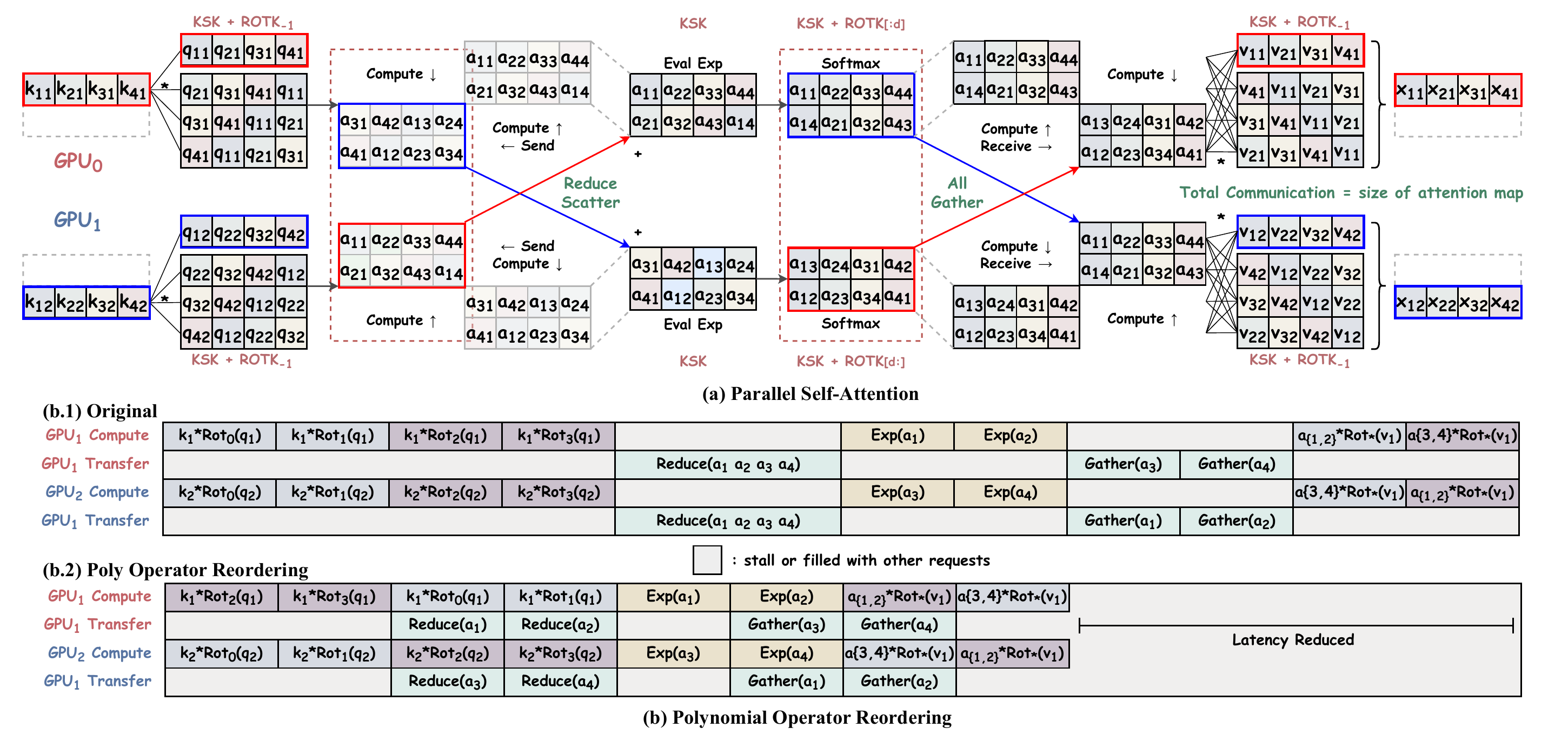}
    \caption{Parallel self-attention: (a) data partition and communication insertion; (b) latency hiding via operator reordering.}
    \Description{Two-panel wide figure for encrypted self-attention. Panel a shows how data is partitioned and where collectives are inserted. Panel b shows operator reordering that overlaps communication with computation to hide latency.}
    \label{fig:encrypted-parallel-attention}
    \vspace{-1em}
\end{figure*}

\subsection{Workload Partition of FFNs}
\label{sec:workload-partition-of-ffns}

For long-sequence encrypted inference, neither encrypted activations nor pre-encoded plaintext weights can be hosted on a single device. In CKKS, weight slices are encoded in the same format as encrypted inputs and participate in accumulation chains determined by the encrypted computation graph. For encrypted activations, slices within a token are accumulation-dependent, while slices across tokens are independent. In contrast, weight slices are structured by \emph{input-output groups}: slices contributing to the same output dimension jointly determine the slots of the output ciphertext.

Unlike plaintext inference, where splitting weights across devices leads to an \texttt{AllGather}, encrypted inference requires \emph{slot-wise accumulation}.
Weight slices contributing to the same output dimension must be accumulated into identical ciphertext slots, resulting in an \texttt{AllReduce}-like behavior.
Since ciphertexts must be transmitted in entirety, partial results cannot be communicated independently. As shown in Figure~\ref{fig:weight-slice-output-dependency}, distributing weight slices from the same output group across devices leads to \emph{partial slot utilization}: each device produces ciphertexts with only a fraction of valid slots.
Communicating such ciphertexts doubles transfer volume without increasing useful work and may additionally introduce redundant nonlinear computation.
More generally, plaintext \texttt{AllGather} patterns along the embedding dimension degenerate into encrypted \texttt{AllReduce} whenever slot utilization is partial.
For example, with 1024 tokens, each linear layer incurs approximately 3.8\,GB of extra inter-GPU communication under such misplacement (Table~\ref{tab:comm-plaintext-vs-encrypted}).

Accordingly, FFN partitioning in \methodname enforces the rule that \emph{weight slices belonging to the same output group must remain co-located}.
Under this placement, at the entry of each linear layer, every device holds $O(T \cdot l / G)$ polynomials corresponding to a disjoint subset of output groups.
Each device initiates a single \texttt{AllGather} on a dedicated communication stream to fetch the required encrypted activations, while simultaneously performing local polynomial multiplication on its resident weight slices. After $G$ rounds of overlapped computation and communication, each device produces its corresponding portion of the output ciphertexts.
The intervening \texttt{GELU} nonlinearity between FFNs is then evaluated locally as an element-wise operation, requiring no additional communication.

\subsection{Workload Partitioning of Self-Attention}
\label{sec:workload-partition-of-self-attn}

Following the placement rules in Section~\ref{sec:application-aware-polynomial-partitioning}, self-attention is partitioned to distribute the scaled dot-product and softmax computation while preserving ciphertext coherence.
After the QKV projections (Section~\ref{sec:workload-partition-of-ffns}), token polynomials remain logically partitioned along the embedding dimension.
Using $2\times$ GPU as an example, token $i$ is split into two polynomial groups, denoted $q/k/v_{i1}$ and $q/k/v_{i2}$, corresponding to disjoint slot regions (Figure~\ref{fig:encrypted-parallel-attention}).

Execution proceeds in three phases.
\emph{(1) Partial attention computation:} each device computes rotate–multiply–accumulate over its local $Q \times K^\intercal$ slices.
Because attention scores aggregate across the full embedding dimension, this phase induces an \texttt{AllReduce}-like dependency at the application level.
\emph{(2) Distributed softmax:} instead of directly performing \texttt{AllReduce}, the intermediate attention polynomials are reshaped via \texttt{ReduceScatter} so that each device holds the data required for local softmax evaluation.
Polynomial approximation of $e^x$ is then applied independently using the numerically stable online softmax procedure~\cite{milakov2018online}, as adopted in prior encrypted attention workloads~\cite{zhang_secure_2025, lu_bumblebee_2023}, without further communication.
\emph{(3) Output aggregation:} the normalized attention scores are \texttt{AllGather}ed so that each device obtains the weights needed to multiply with its local $V$ slices.
The final $A \times V$ computation is then entirely local, producing token-slice outputs without additional synchronization.

During execution, some polynomial slices fully reconstruct tokens while others form partial ciphertexts (see Section~\ref{sec:application-aware-polynomial-partitioning}). The communication operators required to preserve ciphertext correctness in such cases are handled by the rules in Section~\ref{sec:communication-operation-insertion}.

\subsection{Communication Operator Insertion}\label{sec:communication-operation-insertion}

Although the RNS representation causes the system to view all data as uniformly sized polynomials, effective communication placement must respect both application-level aggregation semantics and underlying RNS polynomial dependencies.  
We therefore propose a communication-operator insertion strategy guided jointly by token-wise aggregation patterns and RNS-level data placement.

Following the placement strategy in Section~\ref{sec:application-aware-polynomial-partitioning}, if a set of polynomials residing on the same GPU corresponds to the same token representation, local reduction is first applied before any cross-device communication during \textbf{token-wise aggregation}.  
If all $c_{\text{tok}} \cdot l$ polynomials corresponding to a token are colocated on a single device, no communication is required for embedding-wise aggregation.  
Communication is only inserted when the polynomials corresponding to the same token are distributed across devices; in this case, collectives are introduced exclusively among devices holding the relevant data.  
These collectives follow layer-specific application patterns as described in Section~\ref{sec:workload-partition-of-ffns} and Section~\ref{sec:workload-partition-of-self-attn}.

Additional rules govern communication placement during layer execution.  
For each GPU, if its local polynomial set constitutes a complete ciphertext and the subsequent operation in the computation graph permits reordering, communication is scheduled according to the following rules:  
(1) \emph{Send before bootstrapping}: although the logical workload remains unchanged, communication volume is reduced since bootstrapping restores multiplicative depth and increases the number of limbs $l$.  
(2) \emph{Rescale before send}: rescaling eliminates one limb, directly reducing transfer size.  
(3) \emph{Local reduction before send}: partial reductions are performed locally to minimize communication volume.  

Violating these rules leads to redundant computation and communication: delaying communication until after bootstrapping or rescaling forces devices to transmit enlarged ciphertexts and to recompute equivalent polynomial transformations independently, while skipping local reduction causes multiple devices to perform identical aggregation work on overlapping data. These rules therefore apply uniformly across different layers and applications to avoid redundant computation and unnecessary data movement.



\begin{figure}
    \centering
    \includegraphics[width=\linewidth]{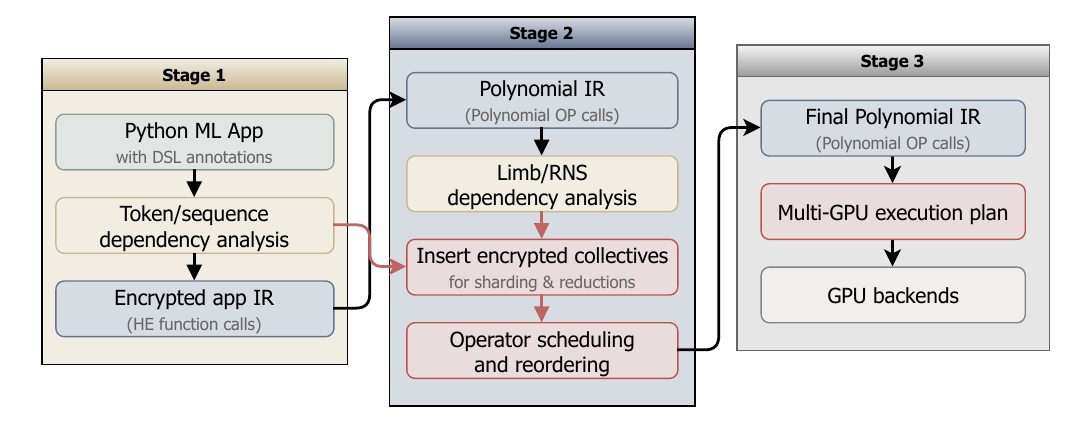}
    \caption{The \methodname compile pass.}
    \Description{Block diagram of the \methodname compile pass, showing lowering from a high-level application representation to encrypted operators, then to polynomial-level instructions, followed by scheduling and multi-GPU execution planning.}
    \label{fig:our-compile-pass}
    \vspace{-1em}
\end{figure}

\subsubsection{Communication Hiding via Operator Reordering}
\label{sec:communication-hiding}

\methodname eliminates GPU stalls caused by embedding-wise aggregation through fine-grained operator reordering.  
When an operation induces embedding wise dependencies (e.g., attention score aggregation), each device maintains two concurrent streams: a compute stream and a communication stream.  
Naively executing the computation in dependency order leads to idle communication streams, since collective operations (e.g., \texttt{ReduceScatter} or \texttt{AllGather}) cannot be issued until all ciphertext components corresponding to a token are ready, causing compute stall.

Figure~\ref{fig:encrypted-parallel-attention}(b) illustrates this inefficiency in parallel self-attention.  
In the original execution order, each device must finish computing all rotated components of a token before communication can begin, leaving the communication stream idle during $QK^\intercal$ computation.  
As a result, communication is serialized after computation, increasing end-to-end latency.

\methodname resolves this inefficiency by reordering polynomial operators across rotation offsets. Instead of computing all rotations for a token contiguously, each device interleaves the computation of different rotation offsets such that partial results for aggregation become available earlier. Specifically, as shown in Figure~\ref{fig:encrypted-parallel-attention}(b), $\text{GPU}_1$ first computes the second half of rotation offsets while $\text{GPU}_2$ computes the first half. This staggered schedule enables the subsequent \texttt{ReduceScatter} for softmax normalization to overlap with ongoing computation on both devices. The same reordering strategy applies to the \texttt{AllGather} following softmax.

By exposing communication opportunities earlier in the execution, \methodname overlaps collective operations with computation, effectively hiding communication latency without introducing additional transfers.

\subsubsection{\methodname Compile Pass}
\label{sec:compile-pass}
HE-based inference systems span multiple abstraction layers, ranging from high-level application graphs to low-level polynomial instructions.
\methodname operates at the intermediate layer between the HE operator control-flow graph (CFG) and the polynomial-instruction level, serving as an orchestration layer for encrypted Transformer inference.

As illustrated in Figure~\ref{fig:our-compile-pass}, the \methodname compile pass proceeds in three stages.
In \textbf{Stage~1}, a Python ML application annotated with a logical data-layout DSL is lowered into an encrypted application IR.
At this stage, \methodname analyzes token- and sequence-level dependencies in the application graph and produces an encrypted IR consisting of HE operator invocations while preserving high-level application semantics.
In \textbf{Stage~2}, the encrypted application is further lowered into a polynomial-level IR.
\methodname analyzes limb- and RNS-level dependencies induced by HE operators and inserts encrypted collective communication primitives to support tensor sharding and encrypted reductions.
Using a hardware-aware cost model that accounts for operator latency and inter-device communication costs, \methodname performs fine-grained operator scheduling and reordering to overlap computation with communication.
Notably, compile-time dependency analysis and collective insertion together take less than 0.7\,s, making this orchestration pass lightweight in practice.
In \textbf{Stage~3}, \methodname generates a deterministic multi-GPU execution plan.
At runtime, each inference request loads its cryptographic keys onto the assigned GPUs, partitions encrypted inputs accordingly, and executes the precomputed schedule in parallel on the target compute backends. This design enables deterministic communication hiding without runtime heuristics or dynamic synchronization.

\subsection{Weak Scaling Analysis}\label{sec:scaling-analysis}

\begin{figure}
    \centering
    \includegraphics[width=1\linewidth]{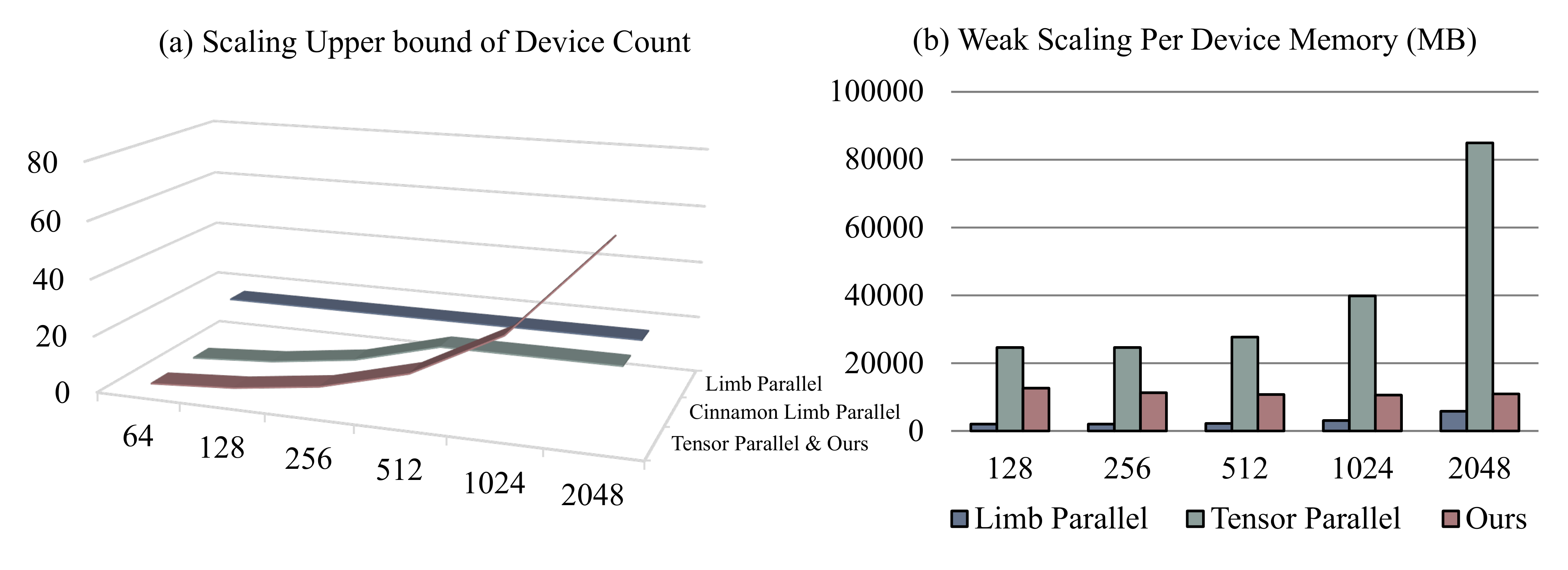}
    \caption{Weak scaling: (a) upper bound on the number of devices as a function of input sequence length (assuming full slot utilization); (b) per-device memory consumption under the maximal feasible device count for each sequence length.}
    \Description{Two-panel weak-scaling figure. Panel a plots the maximum usable device count against input sequence length under full slot utilization. Panel b plots per-device memory at that maximal device count.}
    \label{fig:weak-scaling-bound}
    \vspace{-1em}
\end{figure}


The weak-scaling behavior of existing parallel designs differs fundamentally from that of \methodname. In this section, we analyze weak scaling using the baselines described in Section~\ref{sec:baselines}. State-of-the-art limb-parallel approaches~\cite{al_badawi_multi-gpu_2021, wang_he-booster_2023} are inherently constrained by the number of available RNS limbs at the bootstrapping level as shown in Figure~\ref{fig:weak-scaling-bound}(a). In practice, this limit is typically $l = 14 + 2 = 16$, including two special primes from $P$; since limbs are partitioned across devices, the number of usable GPUs cannot exceed the number of limbs.

Cinnamon~\cite{jayashankar_cinnamon_2025} is further constrained by both limb count and ciphertext count. Although it combines limb-parallel execution with phases in which each device must temporarily hold all components of a ciphertext (e.g., during FFNs), this hybrid strategy restricts scalability to the minimum of the bootstrapping level and the number of concurrently active ciphertexts.

Hydra’s tensor-parallel design~\cite{yang_hydra_2025} scales with sequence length only until memory becomes the bottleneck. As model dimension and sequence length increase, Hydra replicates inputs and stores all QKV ciphertext components on every device. This replication induces substantial memory overhead and leads to per-device memory exhaustion, as shown in Figure~\ref{fig:weak-scaling-bound}(b).

In contrast, \methodname scales with sequence length without inheriting limb-count limits or requiring ciphertext replication. As illustrated in Figure~\ref{fig:weak-scaling-bound}(a,b), \methodname maintains both scalability and memory efficiency, approaching the memory footprint of limb-parallel methods at large sequence lengths while supporting a substantially larger number of devices.

\section{Experiments}

\subsection{Implementation}\label{sec:implementation}

We use LibTorch~\cite{paszke2019pytorchimperativestylehighperformance} as the tensor
container and memory allocator, and employ the CUDA backend from
LiberateFHE~\cite{Liberate_FHE} for all RNS computations on GPU.
Frontend compile pass (see Section~\ref{sec:compile-pass}) is implemented via \texttt{torch.fx} symbolic tracing and operator lowering transformations, which injects the required encrypted operations for a given computation flow. 

We ensure full slot utilization with $s_{\text{tok}} = 64$ and apply BSGS for matrix multiplications. While the Cinnamon baseline relies on hoisting, \methodname disables it by default because hoisting creates intermediate copies proportional to the baby-step size, substantially increasing peak memory usage. Our GPU profiling shows that ciphertext polynomials already saturate the SMs and expose sufficient parallelism, so hoisting provides only marginal throughput gains and does not reduce per-request latency relative to a one-by-one rotate-multiply-accumulate strategy. Therefore, similar to Min-KS in \cite{agrawal_mad_2023}, \methodname adopts the one-by-one approach, saving per-user keys without sacrificing latency.

\textbf{Model.} The end-to-end results in this section are obtained with a full BERT-Base Transformer on the SST-2 task. Under encrypted inference, this model reaches 0.923 accuracy and 0.929 F1 on SST-2. While \methodname's placement and scheduling strategy applies generally to both encoder-style and decoder-style Transformer architectures, we focus the current empirical evaluation on BERT-Base.

\textbf{Encryption Parameters.} We adopt a bootstrappable setting at 128-bit security. The number of slots is $S = 2^{15}$, with polynomial degree $N = 2^{16}$. The RNS modulus is sized at $1661$ bits, providing $|Q_L| = 35$ levels and $|P| = 4$ special primes for bootstrapping.
Following the NEXUS~\cite{zhang_secure_2025} bootstrapping schedule, we set the bootstrapping level to $l_{\text{boot}} = 14$, leaving a usable depth of $L = |Q_L| - l_{\text{boot}} = 21$.

\textbf{Environment.} The 2-GPU experiments are conducted on a platform equipped with an AMD Ryzen Threadripper PRO~7975WX CPU (32~cores), 512\,GB of system memory, and two NVIDIA RTX~A6000 GPUs (48\,GB each) with NVLink P2P connectivity. The 4-GPU experiments are conducted on a server equipped with an AMD EPYC~7763 CPU (64~cores), 2.0\,TiB of system memory, and four NVIDIA A100 GPUs (40\,GB each).

\subsubsection{Baselines}\label{sec:baselines}

We compare \methodname against four state-of-the-art baselines representing distinct parallelization strategies for encrypted and plaintext inference.

\textbf{HEBooster}~\cite{al_badawi_multi-gpu_2021} implements \emph{limb-parallel} execution following prior GPU-based encrypted inference designs.  
Within each GPU, polynomial coefficients are processed in SIMD fashion at the SM level, while ciphertext limbs are partitioned across devices according to the RNS model shown in Figure~\ref{fig:rns-parallelization-model}.

For \textbf{Cinnamon}~\cite{jayashankar_cinnamon_2025}, we implement its output-aggregation key-switching strategy for FFNs.  
Other components of Cinnamon rely on custom hardware modules and specialized instruction sets and are therefore not implemented on GPUs.

As a plaintext tensor-parallel baseline, we implement Megatron-LM–style \textbf{\emph{row-wise} tensor parallelism}~\cite{shoeybi_megatron-lm_2020} across $G$ GPUs, followed by encrypted inference executed independently on each device.  
For \textbf{Hydra}~\cite{yang_hydra_2025}, which adopts \textbf{\emph{column-wise} tensor parallelism}, we replicate the full encrypted input on every GPU, partition FFN weights across devices, and perform all-device aggregation at each layer boundary.  
We further incorporate Hydra’s tree-based task-distribution strategy for nonlinear operators, while their ASIC-specific, CPU-free execution model and switch-based interconnect are not reproducible on GPGPU platforms and are excluded.

These baselines capture complementary pros and cons: \textbf{Limb-parallel methods are memory-efficient} by avoiding replication, but incur heavy communication during key switching. \textbf{Tensor-parallel methods reduce communication rounds} by duplicating inputs or weights, at the cost of increased memory usage. We show that \methodname combines the strengths of both, achieving low communication overhead without sacrificing memory efficiency.

\crpedit{\textbf{Implementation Scope.}
Both Cinnamon and Hydra target ASIC/FPGA platforms with specialized interconnects and unpublished implementations. Since their hardware cannot be replicated on commodity GPUs, our comparison intentionally isolates only the parallelization and task-mapping strategies that can be reproduced on a common GPGPU platform, and does not include ASIC-, ISA-, or interconnect-specific optimizations. This yields an apples-to-apples GPU strategy comparison across baselines and \methodname. All baselines and \methodname incorporate orthogonal and identical compiler optimizations, including late
relinearization and late rescaling~\cite{dathathri_eva_2020}.}

\begin{table*}
\vspace{-1em}
\centering
\caption{Scaling efficiency in terms of latency with different parallel strategies for a fixed 128 input tokens on 2 GPUs. Layer Latency Proportion is a reference value on single device execution.}
\label{tab:layer-strong-scaling-overhead}
\vspace{-1em}
\resizebox{0.95\textwidth}{!}{%
\begin{tabular}{lcc|c|c|c|c|c}
\hline
\multicolumn{1}{c|}{Layer} &
  \multicolumn{1}{c|}{\begin{tabular}[c]{@{}c@{}}Depth\\ on Entry\end{tabular}} &
  \begin{tabular}[c]{@{}c@{}}Layer Latency\\ Proportion (\%)\end{tabular} &
  \begin{tabular}[c]{@{}c@{}}Plaintext Tensor\\ Parallel NVLink\end{tabular} &
  \begin{tabular}[c]{@{}c@{}}HEBooster Limb\\ Parallel NVLink\end{tabular} &
  \begin{tabular}[c]{@{}c@{}}Cinnamon Limb\\ Parallel NVLink\end{tabular} &
  \begin{tabular}[c]{@{}c@{}}Hydra Tensor\\ Parallel NVLink\end{tabular} &
  \begin{tabular}[c]{@{}c@{}}\methodname Parallel\\ NVLink\end{tabular} \\ \hline
\multicolumn{1}{l|}{$Q, K, V$ Linear Projection} &
  \multicolumn{1}{c|}{0} &
  28.78\% &
  49.52\% &
  31.71\% &
  81.71\% &
  89.53\% &
  93.29\% \\
\multicolumn{1}{l|}{$A\gets Q\times K^\intercal / \sqrt{d}$} &
  \multicolumn{1}{c|}{1} &
  16.20\% &
  50.00\% &
  30.75\% &
  30.75\% &
  91.13\% &
  92.37\% \\
\multicolumn{1}{l|}{$A\gets \texttt{Softmax}(A)$} &
  \multicolumn{1}{c|}{2} &
  3.48\% &
  50.00\% &
  30.43\% &
  30.43\% &
  40.52\% &
  89.32\% \\
\multicolumn{1}{l|}{$X\gets A\times V$} &
  \multicolumn{1}{c|}{18} &
  3.29\% &
  50.00\% &
  30.75\% &
  30.75\% &
  91.13\% &
  94.21\% \\
\multicolumn{1}{l|}{$X\gets X\times W_o$} &
  \multicolumn{1}{c|}{19} &
  1.20\% &
  49.87\% &
  34.63\% &
  84.63\% &
  88.32\% &
  94.66\% \\
\multicolumn{1}{l|}{$\texttt{Bootstrapping}$} &
  \multicolumn{1}{c|}{20} &
  1.14\% &
  50.00\% &
  31.50\% &
  61.23\% &
  48.63\% &
  91.32\% \\
\multicolumn{1}{l|}{$X\gets\texttt{LayerNorm}(X)$ (PostNorm)} &
  \multicolumn{1}{c|}{4} &
  2.80\% &
  49.83\% &
  30.11\% &
  30.11\% &
  33.65\% &
  92.32\% \\
\multicolumn{1}{l|}{$\texttt{Bootstrapping}$} &
  \multicolumn{1}{c|}{20} &
  1.14\% &
  50.00\% &
  31.50\% &
  61.23\% &
  48.63\% &
  91.32\% \\
\multicolumn{1}{l|}{$FFN_1:X\gets X\times W_{o1}$} &
  \multicolumn{1}{c|}{4} &
  24.59\% &
  50.00\% &
  34.17\% &
  84.17\% &
  89.33\% &
  93.76\% \\
\multicolumn{1}{l|}{$X\gets\texttt{GeLU}(X)$} &
  \multicolumn{1}{c|}{5} &
  3.04\% &
  50.00\% &
  30.11\% &
  30.11\% &
  47.65\% &
  88.23\% \\
\multicolumn{1}{l|}{$FFN_2: X\gets X\times W_{o2}$} &
  \multicolumn{1}{c|}{19} &
  9.00\% &
  50.00\% &
  31.27\% &
  82.27\% &
  88.35\% &
  93.82\% \\
\multicolumn{1}{l|}{$\texttt{Bootstrapping}$} &
  \multicolumn{1}{c|}{20} &
  1.14\% &
  50.00\% &
  31.50\% &
  61.23\% &
  48.63\% &
  91.32\% \\
\multicolumn{1}{l|}{$X\gets\texttt{LayerNorm}(X)$ (PreNorm)} &
  \multicolumn{1}{c|}{4} &
  3.07\% &
  49.83\% &
  30.11\% &
  30.11\% &
  43.65\% &
  92.32\% \\
\multicolumn{1}{l|}{$\texttt{Bootstrapping}$} &
  \multicolumn{1}{c|}{20} &
  1.14\% &
  50.00\% &
  31.50\% &
  61.23\% &
  48.63\% &
  91.32\% \\ \hline
End to end scaling efficiency &
   &
   &
  \textbf{49.85\%} &
  \textbf{31.86\%} &
  \textbf{53.01\%} &
  \textbf{75.59\%} &
  \textbf{92.98\%} \\ \hline
\end{tabular}%
}\\
\footnotesize{Note: An efficiency of $1.0$ indicates ideal scaling; values around $0.9$ represent high efficiency. For reference, in plaintext settings, SOTA work~\cite{shoeybi_megatron-lm_2020} indicates that tensor parallelism achieves about $95\%$ on 2 GPUs, $82\%$ on 4 GPUs, and $77\%$ on 8 GPUs. Note that while directly applying plaintext parallel outperforms limb parallel in terms of latency, the efficiency in terms of throughput is the worst, with over half of the computation wasted.}
\end{table*}

\subsection{Communication Overhead}

\begin{figure}
    \centering
    \includegraphics[width=\linewidth]{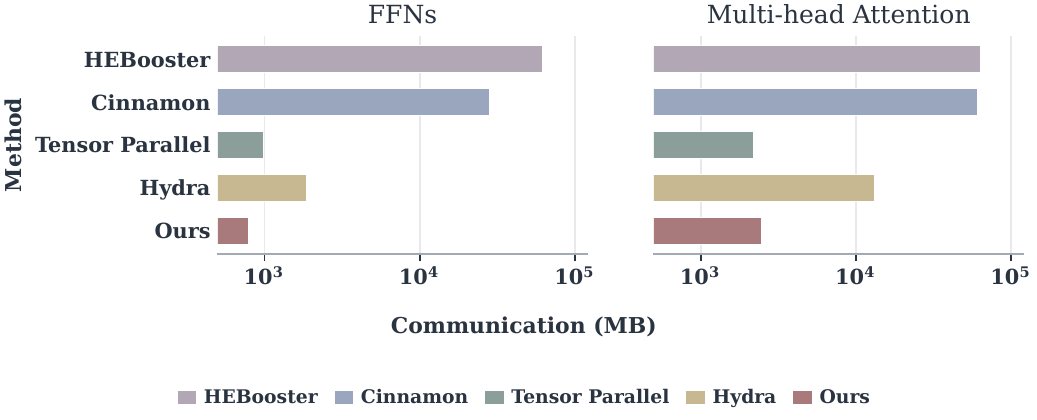}
    \caption{Communication volume on two GPUs with 128 input tokens on BERT.}
    \Description{Bar chart comparing communication volume in megabytes for different methods on two GPUs with 128-token BERT inference. The figure shows much higher communication for limb-parallel baselines than for \methodname.}
    \label{fig:communication-in-mb}
    \vspace{-1em}
\end{figure}

As shown in Figure~\ref{fig:communication-in-mb}, HEBooster exhibits extremely high communication overhead due to frequent key switching.
Both FFNs and self-attention incur communication on the order of $10^5$\,MB, reaching 61{,}735.77\,MB and 63{,}395.76\,MB, respectively.
Cinnamon partially reduces communication in FFNs (to 60{,}550.07\,MB) and during bootstrapping, but still incurs substantial overhead in nonlinear layers and self-attention due to repeated \texttt{CMult} operations.

Directly applying tensor parallelism significantly reduces communication, requiring only three rounds of collectives in FFNs (991.71\,MB).  
However, Hydra’s tensor-parallel design, while avoiding limb-level partitioning, replicates encrypted inputs across devices and performs sequence-wise collectives during self-attention, resulting in 13{,}143.96\,MB of communication.

In contrast, \methodname tightly bounds communication by application-level dependencies.  
Feed-forward layers incur communication that scales as $\Theta(\lceil T/(S/s_{\text{tok}})\rceil \cdot c_{\text{tok}})$, while self-attention requires only a single round of sequence-wise collectives with cost $\Theta((G-1)\cdot T^2/S)$.  
Consequently, \methodname reduces communication to 788.65\,MB in FFNs and 2{,}459.66\,MB in self-attention.  
\crpedit{It isolates the first source of improvement: dependency-aware placement reduces the communicated data itself by matching transfers to application-level dependencies.
The remaining latency gains come from deterministic operator reordering, which overlaps the now-smaller communication with computation rather than further reducing communication volume.
An ablation on two GPUs shows that placement contributes 76.27\% of the speedup at 128 tokens and 88.26\% at 2048 tokens, while reordering contributes the remaining 23.73\% and 11.74\%, respectively.
As sequence length increases, self-attention dominates execution and leaves less latency to hide via overlap, making placement the larger contributor at long sequences.
We quantify these latency-side effects in Section~\ref{sec:benchmarking-scaling-efficiency} and Section~\ref{sec:end-to-end-evaluation}.}

\subsection{Scaling Efficiency of Layer Components}\label{sec:benchmarking-scaling-efficiency}

To analyze the \textbf{efficiency of task-mapping design in terms of latency}, we compare \methodname against plaintext parallelism, HEBooster limb parallelism~\cite{al_badawi_multi-gpu_2021, wang_he-booster_2023}, Cinnamon limb parallelism~\cite{jayashankar_cinnamon_2025}, and Hydra tensor parallelism~\cite{yang_hydra_2025}.  
We evaluate both layer-wise and end-to-end scaling efficiency, computed as $T_1/T_G$ where $T_1$ is the single-GPU latency and $T_G$ is the latency on $G$ GPUs.  
Following prior benchmarks~\cite{park_powerformer_2024, zhang_secure_2025, moon_thor_2024}, we fix the input length to 128 tokens, the maximum length supported by single-device inference, and report results in Table~\ref{tab:layer-strong-scaling-overhead}.

Limb-parallel schemes distribute ciphertext memory evenly across devices, as analyzed in Section~\ref{sec:scaling-analysis}, but nevertheless exhibit \emph{the worst} scaling efficiency.  
In Table~\ref{tab:layer-strong-scaling-overhead}, HEBooster achieves only 31.24\% end-to-end efficiency: the 1617.17s single-GPU latency increases to 2530.62s when distributed, becoming slower than the single-GPU baseline.  
This degradation arises from (1) the massive communication volume during intensive rotation chains, and (2) operation stalls caused by waiting on these transfers.  
We conclude that limb parallelism is suitable for statistical workloads with low key-switching intensity, but fundamentally mismatched to long-sequence encrypted Transformer inference.

Cinnamon improves limb parallelism during \texttt{PMult} matrix multiplications via switching between tensor parallel and limb parallel, yielding higher efficiency for FFNs (about 80\%) and moderate gains for bootstrapping (61\%). However, it still suffers from poor performance on self-attention, where \texttt{CMult} and non-linearities dominate and communication becomes the bottleneck. As a result, Cinnamon reaches only 53.01\% model-wise efficiency.

Hydra tensor parallelism achieves roughly 90\% efficiency on FFNs and the matrix-multiplication segments of self-attention, as these components benefit from replicated inputs and favorable parallel structure.  
However, Hydra’s strategy for nonlinear operations requires transferring ciphertexts based on polynomial-degree ownership, resulting in 40–50\% efficiency in nonlinearities and 48.63\% for bootstrapping.  
Its full-model efficiency reaches 75.87\%.

\methodname achieves the highest overall efficiency across all layers.  
Compared to limb parallelism, \methodname eliminates the dominant communication during key switching via locality-preserving RNS placement, and hides the remaining communication through polynomial-operator reordering.  
Compared to Hydra, \methodname inserts only a single \texttt{ReduceScatter} and a single \texttt{AllGather} per self-attention block, as described in Section~\ref{sec:workload-partition-of-self-attn}.  
Consequently, except for softmax (89.32\%), all layers consistently exceed 90\% efficiency, yielding a 92.98\% end-to-end efficiency, which is \textbf{43.13\% higher than HEBooster, 39.97\% higher than Cinnamon, and 17.11\% higher than Hydra}.

We also observe that layers with identical input sizes may still exhibit different efficiencies due to depth-dependent workload variation. For example, although $FFN_1$ and $FFN_2$ consume the same input shape, they operate at bootstrapping depths 4 and 19, respectively, in Table~\ref{tab:layer-strong-scaling-overhead}. Limb-parallel methods are highly sensitive to multiplicative depth: HEBooster exhibits a 2.90\% discrepancy, and Cinnamon a 1.90\% discrepancy, because communication scales linearly with available depth. In contrast, \methodname is largely insensitive to such variation: its communication-to-compute ratio remains negligible due to depth-invariant polynomial placement and reduced collective frequency.

\subsection{End-to-end Evaluation}\label{sec:end-to-end-evaluation}

\begin{figure}
    \centering
    \includegraphics[width=0.9\linewidth]{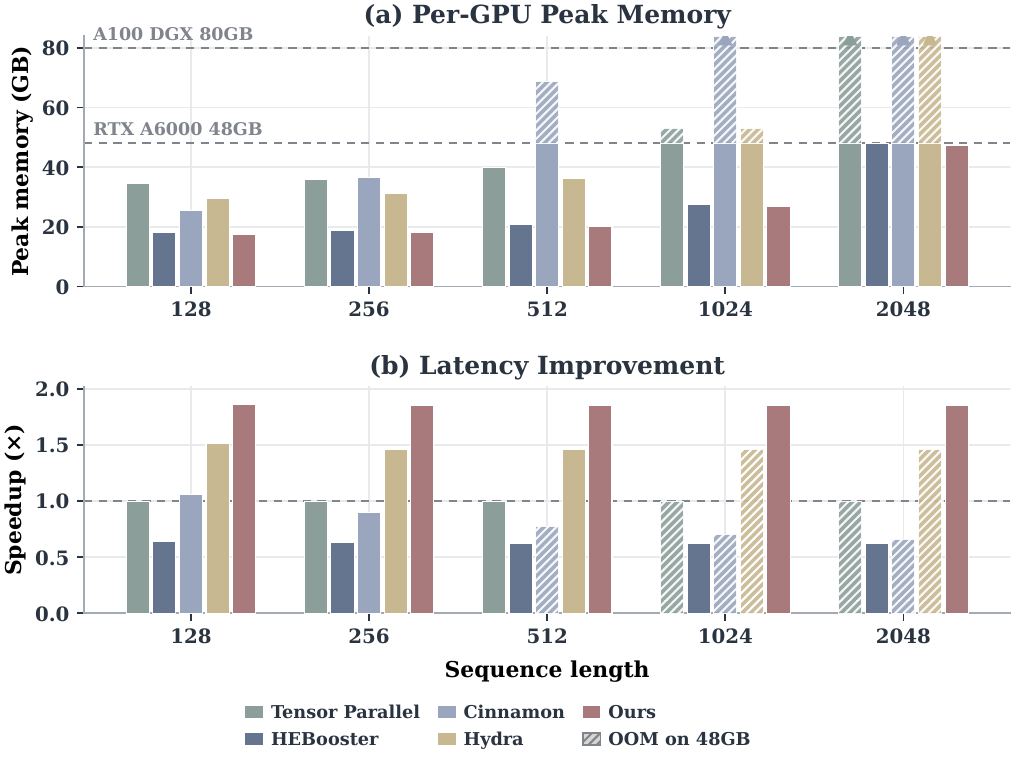}
    \caption{End-to-end inference (a) per-device memory consumption; (b) full-model speed-up, where a baseline $1.0$ corresponds to the latency of processing a full request on a single device. Hatched bar segments and hollow markers denote OOM on 48GB GPU.}
    \Description{Two-panel figure for two-GPU end-to-end inference. Panel a shows per-device memory across input lengths for each method. Panel b shows full-model speed-up relative to single-GPU execution, with hatched or hollow marks indicating out-of-memory cases on a 48 GB GPU.}
    \label{fig:2-gpu-mem-and-speedup}
    \vspace{-1em}
\end{figure}

\crpedit{Beyond layer-level efficiency, we evaluate end-to-end inference performance of the encrypted BERT model on both $2\times$ and $4\times$ GPU systems in Figure~\ref{fig:2-gpu-mem-and-speedup} and Figure~\ref{fig:4-gpu-mem-and-speedup}. In this section, we report per-device memory footprint and end-to-end speedup relative to single-GPU execution as the input length increases from 128 to 2048 tokens.}

\subsubsection{Memory Efficiency}

\crpedit{Figure~\ref{fig:2-gpu-mem-and-speedup}(a) and Figure~\ref{fig:4-gpu-mem-and-speedup}(a) show the same memory ordering across two and four GPUs. HEBooster limb parallelism and \methodname consistently use the least memory because both shard weights and input ciphertexts rather than replicating full activations, whereas tensor-parallel and hoisting-heavy baselines retain a much larger per-device footprint.}

\crpedit{The main difference between the two settings is the OOM boundary, not the trend. Moving from two 48\,GB GPUs to four 40\,GB GPUs reduces \methodname's per-device memory from 17.14--47.15\,GB to 14.55--30.82\,GB, allowing \methodname to remain feasible across the full 2048-token range in both settings. The additional sharding postpones OOM for Hydra and Cinnamon to 512 tokens on four GPUs, but tensor parallelism still overflows at 1024 tokens and both Hydra and Cinnamon still fail at 1024--2048 tokens once replicated activations or hoisted intermediates dominate memory.}

Compared to HEBooster limb parallelism, \methodname uses similar memory with 1.67\%-4.51\% differences along all sequence length. This difference arises because both designs evenly distribute weights and input ciphertexts, while \methodname additionally retaining certain user-specific keys locally on each device. As sequence length increases, the memory footprint of a few key switching keys becomes negligible relative to encrypted activations, causing the memory proportion of the two approaches to converge.

Compared to Cinnamon limb parallelism, \methodname exhibits substantially lower memory consumption.  
Cinnamon performs output-aggregation key switching via hoisting at each layer entry, causing $d$ rotated intermediates derived from the same ciphertext to coexist on every device.  
This design significantly increases memory pressure without providing latency benefits on our platform (see Section~\ref{sec:implementation}).  
Although Cinnamon remains memory-efficient for short sequences (25.32\,GB at 128 tokens, compared to Hydra’s 29.50\,GB), it exhausts device memory at 512 tokens.  
In contrast, \methodname reduces memory usage by 31.59\% at 128 tokens and 50.86\% at 256 tokens.  
As sequence length increases, approaches that replicate intermediate ciphertexts scale poorly, since encrypted activations become the dominant memory component.

Hydra tensor parallelism exhibits a similar limitation.  
At 128 tokens, Hydra consumes 41.27\% more per-device memory than \methodname for a single request, increasing to 45.60\% at 512 tokens.  
This overhead arises because Hydra fully gathers layer outputs at each layer boundary, forcing every device to hold a complete copy of the encrypted activations before the next layer executes.  
As shown in Figure~\ref{fig:memory-breakdown-by-sequence-length}, encrypted activations increasingly dominate memory as sequence length grows.  
These results indicate that efficient long-sequence encrypted inference fundamentally requires distributing activation memory across GPUs, rather than replicating it.

\subsubsection{Performance}

\begin{figure}
    \centering
    \includegraphics[width=0.9\linewidth]{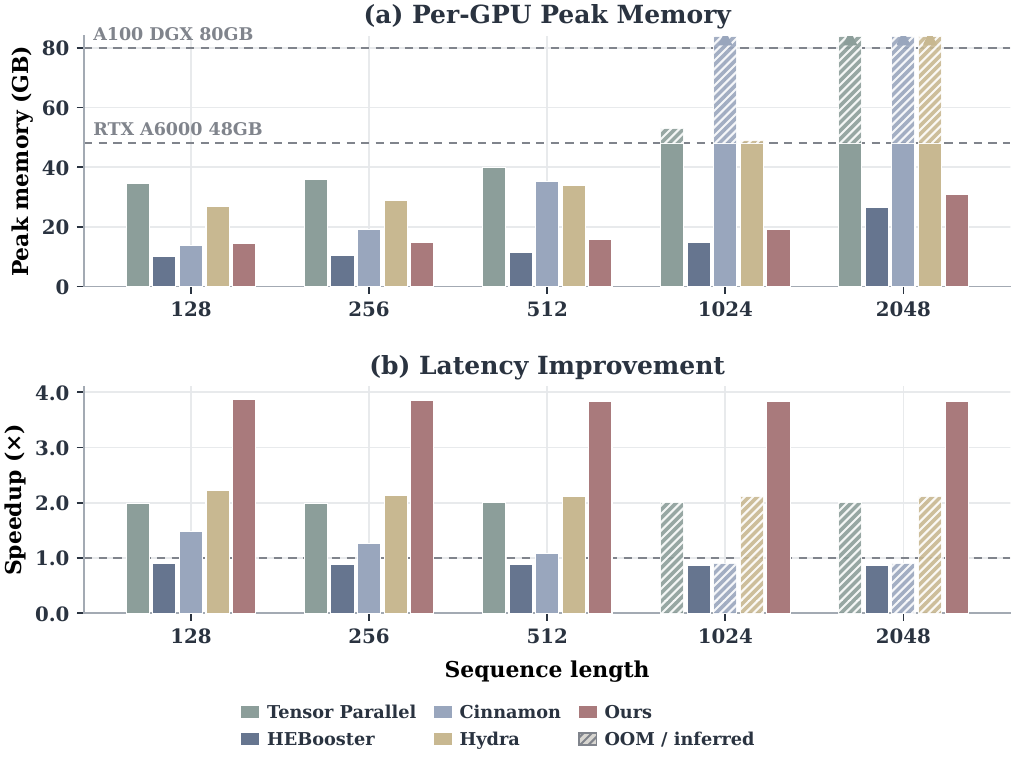}
    \caption{\crpedit{End-to-end inference on four GPUs: (a) per-device memory consumption; (b) full-model speed-up. Hatched entries denote OOM configurations with inferred values used for visualization.}}
    \Description{Two-panel figure for four-GPU end-to-end inference. Panel a shows per-device memory across input lengths for each method. Panel b shows full-model speed-up, with hatched entries marking out-of-memory configurations whose values are inferred for visualization.}
    \label{fig:4-gpu-mem-and-speedup}
    \vspace{-1em}
\end{figure}

\crpedit{Figure~\ref{fig:2-gpu-mem-and-speedup}(b) and Figure~\ref{fig:4-gpu-mem-and-speedup}(b) report the end-to-end speed-up across different sequence lengths. For configurations that exceed device memory, we reuse the same input ciphertexts and weight plaintext slices so the measured speed-up reflects execution behavior rather than repeated data materialization.}
\crpedit{\methodname shows the same trend on two and four GPUs: its speed-up curve remains nearly flat as sequence length grows, reaching 1.850--1.860$\times$ on two GPUs and 3.823--3.865$\times$ on four GPUs. These correspond to scaling factors of about 0.93 and 0.95, indicating that locality-preserving placement removes the dominant communication cost while operator reordering continues to hide the remaining collectives even for long contexts. We emphasize scaling factor as the primary metric because it normalizes hardware-specific throughput differences, but the corresponding end-to-end latencies on the $4\times$A100 platform are 192.8\,s at 128 tokens, 544.2\,s at 512 tokens, and 5036.1\,s at 2048 tokens, following the same overall trend. The four-GPU curve is also slightly higher because those measurements use A100s with faster NVLink than the two-GPU A6000 setup.}

\methodname consistently outperforms all baselines and achieves the highest and most stable speed-up across input lengths. On two GPUs, it reaches a $1.858\times$ speed-up, which is $1.86\times$ faster than directly applying plaintext tensor parallelism, $2.92\times$ faster than HEBooster limb parallelism, $1.75\times$ faster than Cinnamon limb parallelism, and $1.23\times$ faster than Hydra tensor parallelism.

Directly applying plaintext tensor parallelism incurs little communication, but severe slot underutilization and redundant computation reduce its speed-up to only $0.997\times$. This redundancy also causes substantial memory waste, consuming over $2\times$ more per-device memory than HEBooster limb parallelism. HEBooster remains communication-bound and achieves only $0.637\times$ speed-up due to repeated synchronization. Cinnamon improves linear projections and FFNs, yielding a $1.060\times$ speed-up, but continues to lose efficiency once self-attention and nonlinear layers dominate, so its advantage diminishes as FFNs become less dominant at longer sequence lengths.

Hydra comes closest to \methodname with a $1.512\times$ speed-up at 128 tokens, but still lags because it replicates the full $K$ and $V$ and handles nonlinear layers inefficiently. Although Hydra is comparable in the attention score computation ($91.13\%$ vs.\ $92.37\%$ efficiency for $A \gets QK^\intercal / \sqrt{d}$ in Table~\ref{tab:layer-strong-scaling-overhead}), its nonlinearity strategy performs poorly on GPUs, especially in softmax. In particular, Hydra reaches only 40.52\% softmax efficiency compared with 89.32\% for \methodname, so \methodname still delivers an additional $0.346\times$ speed-up while also reducing memory consumption.

\crpedit{The same qualitative ordering holds on four GPUs, but the gap widens once longer sequences stress communication and memory. Tensor parallelism rises only to about $2.0\times$ before OOM at 1024 tokens, Hydra reaches 2.10--2.23$\times$ and Cinnamon 1.08--1.48$\times$ for short inputs but both fail beyond 512 tokens, and HEBooster remains below $1\times$ throughout.}

\section{Related Works}

\textbf{Ciphertext Packing and Application Level Optimizations.} Packing schemes have been extensively studied across diverse HE-based encrypted inference workloads, including CNNs~\cite{al_badawi_towards_2021, kim_secure_2022, aharoni_helayers_2023}, GNNs~\cite{ran_cryptogcn_2022, ran_penguin_2023, kan_ficgcn_2025}, and recent works on Transformers~\cite{park_powerformer_2024, zhang_secure_2025}. These approaches optimize encrypted Transformer inference at the application level, including polynomial approximation for nonlinear functions and algorithmic improvements to encrypted matrix multiplication. These techniques are orthogonal to \methodname, which instead targets cross-operator multi-GPU orchestration under ciphertext/RNS coupling. Meanwhile, recent tensor compilers~\cite{viand_heco_2023, ebel_orion_2025, krastev_tensor_2024} further automate the mapping of tensor operations to HE primitives, simplifying program development and layout management.

\textbf{Compiler Optimization.} Several compiler frameworks focus on parameter selection~\cite{dathathri_chet_2019, reagen_cheetah_2021}, managing multiplicative depth on-the-fly~\cite{dathathri_eva_2020, lee_hecate_2022, lee_performance-aware_2024} and optimizing bootstrapping placement through analytical or learned cost models~\cite{cowan_porcupine_2021, liu_resbm_2025, cheon_dacapo_2024}. These methods are orthogonal to packing-based approaches and primarily target latency reduction and noise management. Not to mention, the resulting dynamic adjustment of modulus-chain length can affect the scalability of RNS under limb-parallel execution, as it changes the number of remaining moduli available for partitioned evaluation.

\textbf{Hardware Acceleration.}
Prior work has explored accelerating individual HE primitives on GPUs~\cite{dai_cuhe_2015, benaissa_tenseal_2021, Liberate_FHE} and on dedicated ASIC and FPGA platforms~\cite{samardzic_f1_2021, kim_bts_2022, agrawal_fab_2023, samardzic_craterlake_2022, yang_poseidon_2023}, demonstrating the feasibility of deploying RLWE-based HE schemes on commodity and specialized hardware.
At the system level, early multi-device designs~\cite{al_badawi_multi-gpu_2021, wang_he-booster_2023} exploit limb- and coefficient-parallelism but incur costly format conversions and scale poorly for rotation-intensive workloads. Application-aware FPGA/ASIC architectures such as Cinnamon and Hydra~\cite{jayashankar_cinnamon_2025, yang_hydra_2025} improve scalability through specialized hardware and interconnects, and are orthogonal to our GPU-focused design. \crpedit{Cerium~\cite{jayashankar_scalable_2025}, also explores GPU acceleration for encrypted inference. Cerium primarily emphasizes kernel- and memory-level optimization within device execution, whereas \methodname focuses on cross-device parallel planning and coherent placement derived from data dependencies across the application and encryption domains.}

\section{Conclusion}

This paper presents \methodname, an application- and encryption-aware multi-GPU execution framework for long-context encrypted Transformer inference. By deriving placement from ciphertext dependencies across both the application and encryption domains, and by reordering operators to overlap the remaining collectives with computation, \methodname reduces communication while avoiding memory overflow and communication-induced stalls. \methodname consistently outperforms prior state-of-the-art designs in both latency and memory efficiency. Our accuracy-validated end-to-end evaluation shows up to 96.62\% scaling efficiency, a 3.86$\times$ end-to-end speedup, and 69.1\% per-device memory reduction on four GPUs. These results indicate that scalable encrypted Transformer inference requires coordinated application-level and encryption-level orchestration, rather than direct adaptation of plaintext parallelism. They also suggest that future HE systems should treat packing, placement, and communication scheduling as a cross-layer optimization problem.

\begin{acks}
This work is partially supported by the National Science Foundation (NSF) under Grants No.~CNS-2348733, No.~CNS-2349538, and No.~CNS-2340777.
\end{acks}

\bibliographystyle{ACM-Reference-Format}
\bibliography{references}


\end{document}